\documentstyle[12pt]{article}

\newcommand{\svline}[1]
{\multicolumn{1}{|c}{#1}}

\newcommand{\EQ}{\begin{equation}}
\newcommand{\EN}{\end{equation}}
\newcommand{\bea}{\begin{eqnarray}}
\newcommand{\ena}{\end{eqnarray}}
\newcommand{\vs}[1]{\vspace{#1 mm}}

\renewcommand{\a}{\alpha}

\def\bbox{{\,\lower0.9pt\vbox{\hrule \hbox{\vrule height 0.2 cm
\hskip 0.2 cm \vrule height 0.2 cm}\hrule}\,}}
\newcommand{\dsl}{\pa \kern-0.5em /}

\newcommand{\pa}{\partial}

\newcommand{\nn}{\nonumber\\}

\newcommand{\lan}{\langle}
\newcommand{\ran}{\rangle} 

\makeatletter
 
 \@addtoreset{equation}{section}
\makeatother

\begin{document}


\topmargin 0pt
\oddsidemargin 0mm
\renewcommand{\thefootnote}{\fnsymbol{footnote}}


\begin{titlepage}

\setcounter{page}{0}
\begin{flushright}
OU-HET 336 \\
hep-th/9912047\\
\end{flushright}

\vs{15}
\begin{center}
{\Large\bf  Instantons  on  Noncommutative 
${ \bf R}^4$  and } \\
\vs{7}
{\Large\bf  Projection Operators } \\
\vs{20}
{\large
Kazuyuki \ Furuuchi\footnote{e-mail address:
furu@het.phys.sci.osaka-u.ac.jp}
}

\vs{10}
{\em %
Department of Physics, Osaka University,
Toyonaka, Osaka 560-0043, Japan} \\
\end{center}

\vs{15}
\centerline{{\large\bf{Abstract}}}
\vs{5}

I carefully study noncommutative
version of ADHM construction of 
instantons, which was
proposed by Nekrasov and Schwarz.
Noncommutative
${\bf R}^4$ 
is described 
as algebra of operators
acting in Fock space.
In ADHM construction of instantons,
one looks for zero-modes
of Dirac-like operator.
The feature
peculiar to 
noncommutative case
is that these zero-modes
project out some states
in Fock space.
The mechanism of 
these projections is clarified
when the gauge group is $U(1)$.
I also construct some
zero-modes when the gauge group is
$U(N)$  
and demonstrate that
the projections also occur, 
and the
mechanism
is similar to the $U(1)$ case.
A physical interpretation
of the projections in 
IIB matrix model
is briefly discussed.

\end{titlepage}
\newpage

\renewcommand{\thefootnote}{\arabic{footnote}}
\setcounter{footnote}{0}
\section{Introduction}

One of the most important
reasons to consider 
physics on noncommutative
spacetime is that 
the behavior 
of the theory at
short distance is  
expected to
become manageable due to the
noncommutativity of the spacetime
coordinates.
It was shown \cite{CDS} that
noncommutative geometry
appear in a definite limit of
string theory, 
BFSS matrix theory \cite{BFSS}
and
IIB matrix theory \cite{IKKT}.
In these cases
noncommutativity should be
relevant to the short scale
physics of D-branes.

Among D-brane systems,
D$p$-brane-D$(p+4)$-brane
bound states
are of interest
because this system
has two different
descriptions:
the one in terms of the
worldvolume theory of
D$p$-branes
and another
in terms of the
worldvolume theory 
of D$(p+4)$-branes.
The D-flat condition
of the worldvolume theory of 
D$p$-brane
coincides with ADHM equations 
\cite{SI}\cite{pinp4},
and D$p$-branes are 
described as instantons in
D$(p+4)$-brane worldvolume theory.
These descriptions should be equivalent
because both describe the
same system, and indeed 
the moduli space of 
the worldvolume theory of
D$p$-branes is
identical to the 
instanton
moduli space.
Turning
constant NS-NS $B$-field 
in the worldvolume of 
D$(p+4)$-branes
causes
noncommutativity
in the worldvolume theory of
D$(p+4)$-branes, and adds 
Fayet-Iliopoulos D-term
in the 
worldvolume theory of
D$p$-branes \cite{DQ}\cite{ABS}.
The equivalence of two
descriptions
follows from the pioneering work of
Nekrasov and Schwarz 
\cite{NS}.
In order to construct
instantons on noncommutative 
${\bf R}^4$,
one adds a constant (corresponding to
the Fayet-Iliopoulos term) to
the ADHM equations.\footnote{%
The case of equivariant 
instanton is studied in \cite{Laza}. }
The modified ADHM equations
describe the
resolutions
of singularities in
the moduli space 
of instantons 
on ${\bf R}^4$ \cite{iNakaj}.
This moduli space has been an important
clue to the 
nonperturbative
aspects of string theory
\cite{IonD}\cite{SumD}\cite{DMVV}%
\cite{Ours}\cite{nakatsu}%
\cite{BPS}
and matrix theory
\cite{ABS}\cite{Dbound}%
\cite{KP}\cite{HilbD}%
\cite{Matso}.
Further studies from 
the viewpoints of
both
string theory and
noncommutative geometry 
were recently
given by \cite{SW}.
More recently
Braden and Nekrasov constructed
instantons on blowups
of ${\bf C}^2$,
which are conjectured to be 
related to  
instantons on 
noncommutative ${\bf R}^4$
\cite{BN}.\footnote{%
In most part of this paper I use the 
term ``commutative" with
usual commutative ${\bf R}^4$ in mind,
and explicitly refer to \cite{BN} when
I compare our noncommutative 
descriptions to
their commutative descriptions.
}

In \cite{NS}
Nekrasov and Schwarz explicitly
constructed some
instanton solutions and
showed that they are
non-singular.
An interesting point is that
they are non-singular
even if their commutative
counterparts in original
ADHM construction are
singular, so-called small
instantons.
In these cases
noncommutativity of the coordinates
actually
eliminates the singular behavior
of the field configurations.
What is special to 
the noncommutative
case
is the appearance of
projection operators which
project out potentially dangerous 
states in Fock space,
where Fock space is introduced to 
describe the 
noncommutative ${\bf R}^4$ \cite{NS}.
The purpose of this paper 
is to investigate this mechanism.
It is shown that this mechanism 
has rich structures,
and gives insight 
in the short scale structures
near the core of instantons on
noncommutative space.
An important point is that
the existence of the projection
forces us to express 
gauge fields in reduced
Fock space where some of the states
have been projected out. 
It is shown that this modification of the Fock space
corresponds to the
modification of the spacetime topology.

An outline of this paper is as
follows.
In section \ref{Pre}, 
gauge theory
on noncommutaive space and
the ADHM construction on
commutative ${\bf R}^4$ are briefly reviewed.
In section \ref{ncADHM}, 
the ADHM construction on noncommutative
${\bf R}^4$ is studied.
The reason
why we must consider the
projections is explained.
In section \ref{secIdeal},
the mechanism of the projections
is clarified
when the gauge group is $U(1)$,
utilizing Nakajima's beautiful
results \cite{iNakaj}\cite{LecNakaj}.
In section \ref{Un}, it is demonstrated
that the similar projections
also occur in $U(N)$ case.
In section \ref{secIIB}, embedding
of the $U(1)$ instanton solution
to IIB matrix model is considered.
The solution is understood
as D-instantons within
D3-brane in IIB matrix model.
It is shown that
the role of the projection is to remove
anti-D-instantons 
and make holes on D3-brane worldvolume.

\ \\
When the previous version of this
paper was at the final stage
of preparation,
the paper \cite{BN}
appeared. 
Some issues discussed in this paper have
commutative counterparts in \cite{BN}.
Explanations on the
modification of spacetime topology
have been added after 
taking into due consideration of
the relation to their 
work.\footnote{%
I would like to thank N. Nekrasov for explaining
their work to me,
and pointing out my misleading
statement in the concluding section 
in the earlier  version
of this paper.}


\section{Preliminaries }\label{Pre}

In this section we briefly review the
theory of
gauge fields on noncommutative ${\bf R}^4$
and ADHM construction
on commutative ${\bf R}^4$,
as preliminaries to
ADHM construction on
noncommutative ${\bf R}^4$.

\subsection{Gauge Fields on
Noncommutative ${\bf R}^4$  }

Noncommutative
${\bf R}^4$ 
is described by an
algebra
generated by
$x^{\mu} \, \,  (\mu  = 1 ,\cdots , 4) $
obeying the commutation relations:
\bea
 \label{noncomx}
[ x^{\mu} ,  x^{\nu}] = i \theta^{\mu\nu} , 
\ena
where $\theta^{\mu\nu}$ is real 
and constant.
In this paper we restrict ourselves
to the case
where $\theta^{\mu\nu}$ is {\em self-dual} and
set\footnote{See \cite{SW} for
the meaning of this choice
of parameters 
in string theory.}
\bea
 \label{theta}
\theta^{12} = \theta^{34} = \frac{\zeta}{4} 
\, \, .
\ena
Then the algebra depends only one constant
parameter $\zeta $.

Introduce the generators of noncommutative
${\bf C}^2 \approx {\bf R}^4$ by 
\bea
z_1 = x_2 + i x_1  , \quad z_2 = x_4 + i x_3 \,  .
\ena
Their commutation relations are:
\bea
 \label{noncom}
 [z_1 , \bar{z}_1] 
=[z_2 , \bar{z}_2] 
= - \frac{\zeta}{2} \quad , \qquad 
\mbox{(others: zero)}.
\ena
We choose $\zeta >  0$ .
The commutation relations (\ref{noncomx})
have a group of automorphisms of the form
$ x^{\mu} \mapsto x^{\mu} + c^{\mu}$,
where $c^{\mu}$ is a commuting real number.
We denote the Lie algebra of this group
by ${\bf \underline{g}}$ .
Following \cite{NS},
we start with 
the algebra End ${\cal H}$ 
of operators acting in the
Fock space
${\cal H} = \sum_{(n_1,n_2) 
\in {\bf Z}_{\geq 0}^2 }
{\bf C} \left| n_1 , n_2    \right\ran$, 
where 
$z, \bar{z}$ are represented
as creation and annihilation operators:
\bea
\sqrt{\frac{2}{\zeta} } z_1 \left| n_1  , n_2    \right\ran
&=& \sqrt{n_1 +1} \left| n_1 + 1 , n_2    \right\ran , \quad
\sqrt{\frac{2}{\zeta} } \bar{z}_1 \left| n_1  , n_2    \right\ran
= \sqrt{n_1 } \left| n_1 -1 , n_2    \right\ran , \nn
\sqrt{\frac{2}{\zeta} } z_2 \left| n_1  , n_2    \right\ran
&=& \sqrt{n_2 +1} \left| n_1 , n_2 +1    \right\ran , \quad
\sqrt{\frac{2}{\zeta} }\bar{z}_2 \left| n_1  , n_2    \right\ran
= \sqrt{n_2} \left| n_1 , n_2 -1    \right\ran .
\ena
The algebra $\mbox{End}\, {\cal H}$ has a 
subalgebra 
of operators which have finite norm;
we define the norm of operators by
$|| a || := 
\mbox{sup}\, || a \phi || / || \phi ||;\,
a \in \mbox{End} {\cal H}, \,
\left| \phi \right\ran \ne 0, \, 
\left| \phi \right\ran \in \mbox{Dom}(a) 
\subset {\cal H}$.
$\mbox{Dom}(a)$ is a domain of operator $a$
and $|| \phi || := 
\left\lan
\phi \right| 
\left. \! \! \phi \right\ran^{1/2}
$.
We denote this algebra by ${\cal A}_\zeta $.
Whenever we consider the derivative
of an operator $a \in {\cal A}_{\zeta}$, 
we assume that it is
also contained in ${\cal A}_{\zeta}$,
i.e. $ \pa_\mu a \in {\cal A}_{\zeta}$
($\pa_{\mu}$ is understood 
as the action of 
${\bf \underline{g}} = {\bf R}^4$ on 
${\cal A}_{\zeta}$ by translation).
The $U(N)$ gauge field
on noncommutative
${\bf R}^4 $ is defined 
as follows.
First
we consider 
$N$-dimensional vector space
${\cal E} := ({\cal A}_{\zeta})^{\oplus N}$ which
carries a right representation 
of ${\cal A}_{\zeta}$:
\bea
{\cal E} \times {\cal A}_{\zeta} 
\ni (e,a) \mapsto ea \in {\cal E} , 
& &e(ab) = (ea)b , \nn
& &e(a+b) = ea + eb, \nn
& &(e + e')a = ea + e'a ,
\ena
for any $e,e' \in {\cal E}$ 
and $ a,b \in {\cal A}_{\zeta}$.\footnote{%
${\cal E}$ is a right module over ${\cal A}_{\zeta}$.
See, for example, \cite{Con}\cite{Landi}.}
The elements of ${\cal E}$ can be thought of
as an $N$-dimensional vector
with
entries in ${\cal A}_{\zeta}$.
Let us consider 
unitary action of
$U$ on the element of ${\cal E}$:
\bea
\label{Ue}
e \rightarrow Ue,
\ena
where $U$ is an $N\times N$ matrix with
its components in ${\cal A}_{\zeta}$,
satisfying $UU^{\dagger}=U^{\dagger}U
= \mbox{Id}_{\cal H}\otimes \mbox{Id}_N$.
$\mbox{Id}_{\cal H}$ is an identity
operator in ${\cal A}_\zeta$ and
$\mbox{Id}_N$ is an $N\times N$ identity matrix.
Under this unitary transformation,
$D e$, the covariant derivative of $e \in {\cal E}$, 
is required to transform covariantly:
\bea
\label{UDe}
 De \rightarrow  UDe.
\ena
The covariant derivative $D$ is
written as
\EQ
D = d+A.
\EN
Here the $U(N)$ gauge field 
$A$ is  
introduced to ensure the covariance,
as explained below. 
$A$ is a matrix valued one-form:
$A = A_{\mu}dx^{\mu}$ with
$A_{\mu}$ being an anti-hermitian
$N\times N$ matrix.
The action of exterior derivative 
$d$ is 
defined as:
\bea
da := (\pa_{\mu} a )\,dx^{\mu}, \quad 
a \in {\cal A}_{\zeta}. 
\ena
$dx^{\mu}$'s commute with $x^{\mu}$
and anti-commute among themselves,
and hence
$d^2 a =0$ for $ a \in {\cal A}_{\zeta}$.
From (\ref{Ue}) and (\ref{UDe}),
the covariant derivative 
transforms as
\bea
 D \rightarrow UDU^{\dagger} .
\ena
Hence the gauge field $A$ 
transforms as
\bea
A \rightarrow UdU^{\dagger} + UAU^{\dagger}.
\ena
The field strength
is defined by
\bea
F := D^2 = dA + A^2.
\ena
We can construct a gauge invariant
action $S$ by\footnote{In this paper
we only consider the case
where the metric 
on ${\bf R}^4$ is flat:
$g_{\mu\nu} = \delta_{\mu\nu}$.}
\bea
S = - \frac{1}{4g^2}\mbox{Tr}_{{\cal H}, U(N)}\,
F_{\mu\nu}F^{\mu\nu}.
\ena

For later purpose,
let us consider a
projection operator 
$P \in {\bf M}_N({\cal A}_{\zeta}), 
P^{\dagger} = P, P^2=P$,
where  ${\bf M}_N({\cal A}_{\zeta})$ denotes
the algebra of $N \times N$
matrices with their entries in ${\cal A}_{\zeta}$.
For every projection operator $P$,
we can consider
vector space $P {\cal E}$:\footnote{%
$P {\cal E}$ is a right projective module
over ${\cal A}_{\zeta}$.
}
\bea
e \in P{\cal E} \Longleftrightarrow
e \in {\cal E} , \, e = P e
\ena
We can consider unitary action on $P{\cal E}$:
\bea
 e \rightarrow U_{\scriptscriptstyle P} e,  
\, \,  & &U_{\scriptscriptstyle P} 
 = P U_{\scriptscriptstyle P} = U_{\scriptscriptstyle P} P, \nn
& &U_{\scriptscriptstyle P}^{\dagger }  
U_{\scriptscriptstyle P} =
U_{\scriptscriptstyle P} 
U_{\scriptscriptstyle P}^{\dagger }
= P.
\ena
We can  construct 
covariant derivative $D_{\scriptscriptstyle P}$ for
$P {\cal E}$ by
\bea
 \label{Pd1}
D_{\scriptscriptstyle P} = Pd + A, \quad A = PA = AP.
\ena
Notice that $D_{\scriptscriptstyle P} 
= PD_{\scriptscriptstyle P}$.
We require $D_{\scriptscriptstyle P} e$ to transform
as 
\bea
D_{\scriptscriptstyle P} e \rightarrow 
U_{\scriptscriptstyle P} D_{\scriptscriptstyle P} e .
\ena
Then the covariant derivative
$D_{\scriptscriptstyle P}$ must transform as
\bea
 \label{PUPDP}
D_{\scriptscriptstyle P} \rightarrow 
 U_{\scriptscriptstyle P} D_{\scriptscriptstyle P} 
U_{\scriptscriptstyle P}^{\dagger}.
\ena
For any $ e \in P{\cal E}$, one can show
\bea
U_{\scriptscriptstyle P} D_{\scriptscriptstyle P}
 U^{\dagger}_{\scriptscriptstyle P} e
&=&  U_{\scriptscriptstyle P} (Pd + A)  
U^{\dagger}_{\scriptscriptstyle P} e
 =   U_{\scriptscriptstyle P}d 
(P(U^{\dagger}_{\scriptscriptstyle P} e))
 + U_{\scriptscriptstyle P} A
 U^{\dagger}_{\scriptscriptstyle P} e \nn
&=&  U_{\scriptscriptstyle P} 
PdU^{\dagger}_{\scriptscriptstyle P} e
+  PU_{\scriptscriptstyle P} 
(U^{\dagger}_{\scriptscriptstyle P} de)
    + U_{\scriptscriptstyle P} A 
U^{\dagger}_{\scriptscriptstyle P} e
    \quad (U_{\scriptscriptstyle P}P 
=PU_{\scriptscriptstyle P})\nn
&=& P de + (U_{\scriptscriptstyle P} 
dU^{\dagger}_{\scriptscriptstyle
  P} + U_{\scriptscriptstyle P} A 
U^{\dagger}_{\scriptscriptstyle P}) e.
\ena
Hence the gauge transformation rule of the
gauge field
$A$ is given by
\bea
A \rightarrow U_{\scriptscriptstyle P} 
dU^{\dagger}_{\scriptscriptstyle P} 
+ U_{\scriptscriptstyle P} A 
U^{\dagger}_{\scriptscriptstyle P}.
\ena
The field strength becomes
\bea
 \label{PF1}
F &:=& D_{\scriptscriptstyle P}^2  \nn
  &=& PdA + A^2 + PdPdP .
\ena
Indeed, for $e \in P{\cal E}$, one can show
\bea
 \label{FeP}
Fe &=& (Pd + A)(Pde + Ae) \nn
   &=& Pd(Pde) + Pd(Ae) + APde + A^2 e \nn
   &=& Pd(Pde) + PdAe + A^2e,
\ena
and since $e = Pe$ and $P^2 = P$, we can
rewrite (\ref{FeP}) using following equations:
\bea
Pd(Pde) &=& Pd(Pd(Pe)) \nn
        &=& Pd(PdPe + Pde) \nn
        &=& PdPdPe-PdPde+PdPde \nn
        &=& PdPdP e.
\ena
Hence we obtain (\ref{PF1}).
We can construct a gauge invariant
action $S_{\scriptscriptstyle P}$ by
\bea
S_{\scriptscriptstyle P} = 
-\frac{1}{4g^2}\mbox{Tr}_{{\cal H}, U(N)}\,
PF_{\mu\nu}F^{\mu\nu}P.
\ena

Gauge field $A$ is 
called anti-self-dual, 
or instanton, if its field strength satisfies
the conditions:
\bea
 \label{self}
F^+ := 
\frac{1}{2}(F + * F ) = 0 ,
\ena
where $*$ is the Hodge star.


\subsection{Review of ADHM Construction
on Commutative ${\bf R}^4$ }

ADHM construction 
\cite{ADHMconst}
is the way to obtain
anti-self-dual gauge field on ${\bf R}^4$
from solutions
of some quadratic matrix equations.
More specifically,
in order to construct
anti-self-dual $U(N)$ gauge field
with instanton number $k$ ,
one starts
from the following data (ADHM data):
\begin{enumerate}
 \item A pair of complex hermitian vector spaces
       $V =  {\bf C}^k $ and $W = {\bf C}^N $ . 
 \item The operators 
     $B_1 , B_2 \in Hom(V,V) ,
      I \in Hom(W,V) , J = Hom(V,W) $\ %
  satisfying the equations 
\bea
 \label{ADHM}
\mu_{\bf R}
 &=& [B_1 , B_1^{\dagger}] +  [B_2 , B_2^{\dagger}] 
        + II^{\dagger} - J^{\dagger} J = 0 , \nn  
\mu_{\bf C} &=& [B_1 , B_2 ] + IJ = 0 .
\ena  
\end{enumerate}
Next 
define Dirac-like operator 
${\cal D }_{z} :
V \oplus V \oplus W \rightarrow V \oplus V $ by
\bea
 \label{Dz}
& &{\cal D}_z = 
\left(
 \begin{array}{c}
  \tau_z \\
  \sigma_z^{\dagger }
 \end{array}
\right)  , \nn
& &\tau_z =
(\, B_2 - z_2 ,\, B_1 - z_1 , \, I \, ) , \nn
& &\sigma_z^{\dagger} =
(\, - (B_1^{\dagger} -\bar{z}_1) , \,
        B_2^{\dagger} - \bar{z}_2  , \, J^{\dagger} \, ) .
\ena
(\ref{ADHM})
is equivalent to the set of equations
\bea
\label{key}
\tau_z \tau_z^{\dagger}
=\sigma_z^{\dagger} \sigma_z , \quad
\tau_z \sigma_z = 0 ,
\ena
which are important conditions
in ADHM
construction.
There are $N$ zero-modes of 
${\cal D}_z$ :
\bea
 \label{czerom}
{\cal D}_z \psi^{(a)} = 0 ,
\quad a= 1, \ldots , N.
\ena
We can choose orthonormal basis
of the space of zero-modes:
\bea 
\psi^{(a)\dagger } \psi^{(b)} = 
\delta^{ab}.
\ena
The change of basis in the
space of orthonormalized zero-modes
$\psi^{(a)}$ becomes
$U(N)$ gauge symmetry.
Anti-self-dual $U(N)$ gauge
field is constructed 
by the formula
\bea
\label{Amu}
A^{ab} 
= \psi^{(a)\dagger } d \psi^{(b)} .
\ena
There is an action of $U(k)$ that does not 
change (\ref{Amu}) :
\bea
 \label{Uk}
(B_1, B_2 ,I , J ) \longmapsto
(g B_1 g^{-1} , g B_2  g^{-1} , g I , J g^{-1} ) , 
 \qquad g \in U(k) .
\ena
The moduli space of anti-self-dual
$U(N)$ gauge field  
with instanton number $k$ is given by
\bea
  \label{moduli}
{\cal M} (k,N)
 = 
\mu^{-1}_{\bf R}(0) \cap \mu^{-1}_{\bf C}(0) /U(k) ,
\ena
where the action of $U(k)$ is the one
given in (\ref{Uk}). 
When $(B_1, B_2 ,I , J )$ is a fixed point
of $U(k)$ action,
${\cal M} (k,N)$ is singular.
Such a singularity corresponds
to an instanton shrinking to zero size.


\section{ADHM Construction
on Noncommutative ${\bf R}^4$ and
the Appearance of Projection Operator}\label{ncADHM}

The singularities in (\ref{moduli}) has 
a natural resolution \cite{iNakaj}.
Modify
(\ref{ADHM}) to 
\bea
 \label{ADHMzeta}
\mu_{\bf R}
 &=& [B_1 , B_1^{\dagger}] 
+  [B_2 , B_2^{\dagger}] 
        + II^{\dagger} - J^{\dagger} J 
= \zeta \, \mbox{Id}_V 
,  \nn
\mu_{\bf C} &=& [B_1 , B_2 ] + IJ = 0 ,
\ena  
and consider the space
\bea
 \label{Mzeta}
{\cal M}_{\zeta } (k,N) =
\mu_{\bf R }^{-1}(\zeta \, \mbox{Id}_V ) \,   
\, \cap \mu_{\bf C}^{-1} (0) \, 
/ U(k) . 
\ena
Then ${\cal M}_{\zeta } (k,N) $ is a smooth 
$4kN$ dimensional 
hyper-K\"ahler manifold.
Although the absence of singularities is 
interesting from
the physical point of view, 
construction of instantons 
from
(\ref{ADHMzeta})
does not work straightforwardly.
The main obstruction is that
the key equations in (\ref{key}) 
are not satisfied on the usual 
commutative ${\bf R}^4$.
However, 
Nekrasov and Schwarz
noticed that 
$\tau_z$ and $\sigma_z$
do satisfy (\ref{key}) if the 
coordinates are noncommutative
as in
(\ref{noncom}) \cite{NS}.
Once (\ref{key}) is satisfied,
we can expect that the
construction of instantons is
similar to the usual commutative case.
But there are some features
peculiar to the 
noncommutative case.
Especially,
since the ADHM construction 
on noncommutative ${\bf R}^4$
starts
from (\ref{Mzeta}) where
small instanton singularities
have been resolved,
one expects that 
crucial difference 
will appear
when the size of the instanton
is small.
It is interesting to study
such situations
and see how
the effects of noncommutativity
appear.

The ADHM construction 
on noncommutative 
${\bf R}^4$ 
is as follows
\cite{NS}.
We define operator
${\cal D}_z :
(V \oplus V \oplus W ) \otimes {\cal A}_{\zeta}
\rightarrow ( V \oplus V ) \otimes {\cal A}_{\zeta} $ 
by the same formula (\ref{Dz}):
\bea
& &{\cal D}_z = 
\left(
 \begin{array}{c}
   \tau_z \\
   \sigma_z^{\dagger }
 \end{array}
\right) , \nn
& & \tau_z =
(\, B_2 - z_2 ,\, B_1 - z_1 , \, I \, ), \nn
& & \sigma_z^{\dagger}  =
( \, - (B_1^{\dagger}-\bar{z}_1) , \,
  B_2^{\dagger} - \bar{z}_2  , \, J^{\dagger} \, ) .
\ena
The operator 
${\cal D}_z {\cal D}^{\dagger }_z :
(V \oplus V) \otimes {\cal A}_{\zeta}
\rightarrow
(V \oplus V) \otimes {\cal A}_{\zeta}$
has a block diagonal form
\bea
 \label{box}
{\cal D}_z {\cal D}_z^{\dagger } 
=
\left(
 \begin{array}{cc}
  \Box_z & 0   \\
     0   & \Box_z 
  \end{array} 
 \right), \qquad
\Box_z \equiv \tau_z \tau_z^{\dagger}
       = \sigma_z^{\dagger } \sigma_z  
\ena
which is a consequence of (\ref{key})
and important for ADHM construction.
Next we look for solutions to the 
equation 
\bea
 \label{zeroPsi}
{\cal D}_z \Psi^{(a)} = 0  \quad
( a = 1, \ldots , N ),
\ena
where the components of $\Psi^{(a)} $ are 
{\em operators}: \ %
$\Psi^{(a)} : 
{\cal A}_{\zeta} 
\rightarrow (V \oplus V \oplus W )\otimes {\cal A}_{\zeta} $.
If we can normalize
$\Psi^{(a)} $'s as 
\bea
 \label{norm}
\Psi^{\dagger (a)} \Psi^{(b)}  
\stackrel{?}{=} 
\delta^{ab}\,  \mbox{Id}_{\cal H}  ,
\ena 
we can construct anti-self-dual
$U(N)$ gauge field by the same 
formula (\ref{Amu}):
\bea
A^{ab}
 \label{ncA} 
= \Psi^{(a) \dagger } d \Psi^{(b)} , 
\ena
where $a$ and $b$ are $U(N)$ indices.
Then the field strength becomes
\bea
 \label{FS}
F &=&F^-_{\mbox{\tiny ADHM}} 
\equiv
\Psi^{\dagger}
\left(
d \tau_z^{\dagger} 
\frac{1}{\, \, \Box_z } d \tau_z
+
d \sigma_z 
\frac{1}{\, \, \Box_z } d \sigma_z^{\dagger}
\right)
\Psi
\nn
&=&\! \! \! \! \! \!
\begin{array}{ccc}
\bigl(\, \psi_1^{\dagger} \,& \,
 \psi_2^{\dagger} \,& \, \xi^{\dagger} \, 
\bigr) \\
   &  &  \\
   &  &  
\end{array} \! \!
\left(
 \begin{array}{ccc}
dz_1 \frac{1}{\, \, \Box_z}d\bar{z_1} 
+ d\bar{z_2} \frac{1}{\, \, \Box_z} dz_2   & 
  -dz_1 \frac{1}{\, \, \Box_z} d\bar{z_2} 
       + d\bar{z_2}\frac{1}{\, \, \Box_z} 
dz_1 & 0 \\
-dz_2 \frac{1}{\, \, \Box_z} d\bar{z_1} 
+ d\bar{z_1}\frac{1}{\, \, \Box_z} dz_2 & 
  dz_2 \frac{1}{\, \, \Box_z} d\bar{z_2} 
  + d\bar{z_1} \frac{1}{\, \, \Box_z} dz_1 & 0 \\
 0 & 0 & 0
 \end{array}
\right) 
\left(
\begin{array}{c}
\psi_1 \\
\psi_2 \\
\xi
\end{array}
\right) \nn
\quad
\ena
where we have written
\bea
& &\Psi \equiv
\left(
\begin{array}{c}
\psi_1 \\
\psi_2 \\
\xi
\end{array}
\right) \equiv
\left(
\begin{array}{ccc}
 & & \\
\Psi^{(1)} & \cdots & \Psi^{(N)} \\
 & &
\end{array}
\right) , \qquad
\begin{array}{c}
\psi_1 : {\bf C}^N \otimes{\cal A}_{\zeta} 
        \rightarrow V \otimes {\cal A}_{\zeta} , \\
\psi_2 : {\bf C}^N  \otimes{\cal A}_{\zeta} 
        \rightarrow V \otimes {\cal A}_{\zeta} ,  \\
\, \,  \xi \, \,  : 
{\bf C}^N  \otimes{\cal A}_{\zeta} 
      \rightarrow W \otimes {\cal A}_{\zeta}. 
\nn
\end{array}
\ena
The derivation is similar
to the commutative case.
The field strength in (\ref{FS}) is 
anti-self-dual.

However, 
as we will see shortly,
there are some states in
${\cal H}$  which are annihilated by
$\Psi^{(a)} $ for some $a$.
More precisely, all the
components
of $\Psi^{(a)} $ annihilate
those states.
This is not a special phenomenon,
and the study of this
phenomenon
is the purpose of this paper.
Let us consider the case
where there is one
such zero-mode 
$\Psi^{(1)} $.
In that case we cannot 
normalize 
$\Psi^{(1)} $
as in (\ref{norm}).  
We may normalize
$\Psi^{(1)}  $ as
\bea
 \label{normP}
  \Psi^{(1)\dagger } \Psi^{(1)} 
=P  ,
\ena
where $P \in {\cal A}_\zeta$ is a projection operator 
that projects out the states annihilated by 
$\Psi^{(1)} $.
However,
the projection operator
gives additional 
contribution to the field strength,
because the projection operator
depends on $z$ and $\bar{z}$.\footnote{%
For example, 
$\left|0,0\right\ran
\left\lan0,0\right| =
: e^{-\frac{2}{\zeta} 
(z_1\bar{z}_1+z_2\bar{z}_2) } : $ ,
where : : means normal ordering.
}  
The derivative of the projection
operator gives additional contribution
to the field strength,
which is not anti-self-dual.

The appearance of the
projection operator
$P$ indicates that we should
consider restricted vector space
$P {\cal E}$ rather than ${\cal E}$.
Indeed, as we will see shortly,
ADHM construction
perfectly works in this setting.

Let us 
concentrate on the
simplest $U(1)$ case.
The covariant derivative is given by the
formula (\ref{Pd1}):
\bea
 \label{Pd}
D_{\scriptscriptstyle P} = Pd + A,
\ena
with $A=PAP$.
The field strength is given by (\ref{PF1}):
\bea
 \label{PF}
F = PdA + A^2 + PdPdP.
\ena
We can construct anti-self-dual gauge field
by putting
\bea
 \label{PAP}
A= \Psi^{\dagger} d \Psi P,
\ena
where $\Psi$ is a zero-mode
of ${\cal D}_z$ and normalized as
$\Psi^{\dagger} \Psi = P$. 
Note that $\Psi^{\dagger} = P \Psi^{\dagger}$.
Let us check that (\ref{PAP}) is really
anti-self-dual.
The first term in (\ref{PF}) becomes 
\bea
PdA = 
Pd\Psi^{\dagger}d\Psi P -
P \Psi^{\dagger} d\Psi dP.
\ena
The last term above can be 
rewritten as
\bea
 \label{pdpdp}
P \Psi^{\dagger} d\Psi dP
&=& P ( d (\Psi^{\dagger} \Psi ) 
      - d\Psi^{\dagger}\Psi ) dP \nn
&=& PdPdP - P d\Psi^{\dagger}\Psi  dP.
\ena
The first term in (\ref{pdpdp})
cancels $PdPdP$ in (\ref{PF}).
The last term in (\ref{pdpdp})
vanishes when acting on $e = P e \in P{\cal E}$,
since $\Psi  dP P = - \Psi P d(1-P) P = 0$.
The second term in (\ref{PF}) becomes
\bea
 \label{A2}
A^2 &=& 
P \Psi^{\dagger} d\Psi \Psi^{\dagger} d\Psi P \nn
&=&
P (d (\Psi^{\dagger} \Psi) ) - d\Psi^{\dagger}\Psi)
\Psi^{\dagger} d\Psi P    \nn
&=&
PdP\Psi^{\dagger} d\Psi 
- P d\Psi^{\dagger} \Psi \Psi^{\dagger}   d\Psi P.
\ena
The first term in (\ref{A2}) vanishes
because
$PdP\Psi^{\dagger} 
= -Pd(1-P)P\Psi^{\dagger} = 0$.
Then the field strength becomes
\bea
 \label{FP}
F 
&=& Pd\Psi^{\dagger}
      (1-\Psi\Psi^{\dagger}) d\Psi P \nn
&=& P F^-_{\mbox{\tiny ADHM}} P    = F^-_{\mbox{\tiny ADHM}}, 
\ena
where $F^-_{\mbox{\tiny ADHM}}$ is defined in (\ref{FS}) and
is anti-self-dual.
Generalization to $U(N)$ case is straightforward.

The absence of the singular behavior
in the field configuration follows
rather straightforwardly from the explicit
formula (\ref{FS}).
Since we have normalized the zero-modes
in the subspace
where zero-modes do not vanish, these normalized
zero-modes are well defined.
Moreover, as shown in
appendix \ref{A}, the operator
$\Box_z$ has no zero-mode
and hence its inverse
does not cause divergences.
Therefore from the explicit formula (\ref{FS}),
we cannot see any 
source of divergences
in (\ref{FS}) or (\ref{FP}), either.


\section{$U(1)$ Instantons and Projection Operators}\label{secIdeal}

\subsection{%
Projection Operators in
$U(1)$ Instanton Solutions and 
Relation to the Ideal}

In the previous section
it is shown how to construct
anti-self-dual
gauge field when
the zero-mode annihilates some states.
Then the natural question is:
``How the 
states annihilated 
by the zero-modes should be determined? "
In this section
the answer to this question is given when 
the gauge group is $U(1)$.

Let us consider the solution to the equation
\bea
 \label{zeroV}
{\cal D}_z \left| {\cal U} \right\ran = 0 , 
\ena
where 
$ \left| {\cal U} \right\ran 
\in {\cal H}^{\oplus k}  \oplus
    {\cal H}^{\oplus k}  \oplus  
    {\cal H}$, 
i.e.
the components of 
$ \left| {\cal U} \right\ran $
are {\it vectors} 
in the Fock space ${\cal H} $.
We call $ \left| {\cal U} \right\ran $
``vector zero-mode" and
call $\Psi $ 
in (\ref{zeroPsi}) 
``operator zero-mode".
We can construct
operator zero-mode
if we know all the vector zero-modes.
The advantage of considering
vector zero-modes
is that we can relate
them to
the ideal discussed in 
\cite{Nakaj}\cite{LecNakaj}.
The point is that we can
regard vector zero-modes
as holomorphic vector bundle 
described
in purely commutative
terms.
Noncommutativity
appears when we construct
operator zero-mode 
treating
all the vector zero-modes
as a whole.

Let us write
\bea
\label{veczero}
\left| {\cal U} \right\ran = 
\left(
 \begin{array}{c}
 \left| u_1 \right\ran \\
  \left| u_2 \right\ran \\
  \left|\, f \, \right\ran 
\end{array}
\right) ,
\qquad
 \begin{array}{l}
 \left| u_1 \right\ran \equiv
   u_1(z_1,z_2) \left|\,  0,0 \,\right\ran \\
\left| u_2 \right\ran \equiv
  u_2(z_1,z_2)\left|\, 0,0\, \right\ran \\
  \left|\, f \, \right\ran 
\equiv  f (z_1,z_2)\left|\, 0,0\, \right\ran 
\end{array}
\ena
where
$
\left| u_1 \right\ran ,
\left| u_2 \right\ran  \in 
 {\cal H}^{\oplus k}$
i.e. they are
vectors in $V = {\bf C}^k$ 
and vectors in ${\cal H}$,
and
$
\left|\, f \, \right\ran \in {\cal H}
$.
The space of the solutions 
of (\ref{zeroV}) , i.e.
$
\ker {\cal D}_z =
\ker \tau_z \cap
 \ker \sigma_z^{\dagger } 
\simeq 
\ker \tau_z/ \mbox{Im}\, \sigma_z $\footnote{%
Notice that 
since
$ \tau_z \sigma_z = 0$ ((\ref{key})),
$\ker \tau_z/ \mbox{Im}\, \sigma_z $ is 
well defined. 
$\ker \tau_z \cap
 \ker \sigma_z^{\dagger } 
\simeq 
\ker \tau_z/ \mbox{Im}\, \sigma_z $ is understood
as follows: the condition
$\ker \sigma_z^{\dagger }
\left|\, {\cal U} \, \right\ran  = 0$ fixes
the ``gauge freedom" 
mod Im $\sigma_z$
in $\ker \tau_z/ \mbox{Im}\, \sigma_z $. }
is isomorphic to the ideal ${\cal I}$ 
defined by 
\bea
 \label{ideal}
{\cal I} 
= \Bigl\{\, f(z_1,z_2) \, 
\Bigm|\, \, f(B_1, B_2) = 0 \, \, \Bigr\} ,
\ena
where $B_1$ and $B_2$ together with $I$ and $J$ give
a solution to (\ref{ADHMzeta}).
In $U(1)$ case, one can show 
$J = 0$, and
the isomorphism is given by the inclusion 
of the third factor in (\ref{veczero}) 
\cite{Nakaj}\cite{LecNakaj}.
\bea
\label{inclusion}
\ker \tau_z/ \mbox{Im}\, \sigma_z\,
 \hookrightarrow {\cal O}_{{\bf C}^2}\,:\quad
\left| {\cal U} \right\ran = 
\left(
 \begin{array}{c}
 \left| u_1 \right\ran \\
  \left| u_2 \right\ran \\
  \left|\, f \, \right\ran 
\end{array}
\right)\,  \hookrightarrow f(z_1,z_2).
\ena
Let us define  ``ideal state" by 
\bea
\left| \varphi \right\ran 
\in \mbox{ideal states of ${\cal I}$ }
\Longleftrightarrow
\exists f( z_1, z_2 ) \in {\cal I} , \quad
\left| \varphi \right\ran =
f( z_1, z_2 ) \left| 0,0 \right\ran ,
\ena
and denote the 
space of all the ideal states by
${\cal H}_{{\cal I} }$.
We define ${\cal H}_{/{\cal I} }$
\footnote{The meaning of this notation
is as follows:
${\cal H}_{/{\cal I} }$ corresponds to
${\bf C}[z_1,z_2]/ {\cal I }$, where
${\bf C}[z_1,z_2]$ is the ring of polynomials
of $z_1 , z_2$.}
as a 
subspace in ${\cal H}$
orthogonal to ${\cal H}_{{\cal I} }$ :
\bea
\left| g \right\ran \in {\cal H}_{/{\cal I} }\,
\Longleftrightarrow  
\forall f(z_1, z_2) \in {\cal I} , \quad
\left\lan 0 \right|
f^{\dagger}( \bar{z}_1, \bar{z}_2 ) 
\left| g \right\ran = 0 .
\ena
${\cal H}_{/{\cal I}}$ is a $k$ dimensional
space \cite{LecNakaj}.
Let us denote the 
complete basis of
${\cal H}_{/{\cal I}}$ by
$\left|\, g_{\a}  \, \right\ran , \a=1, 2,\cdots ,k$, 
and the orthonormalized
complete basis of ${\cal H}_{{\cal I}}$ by 
$  \left|\, f_i \, \right\ran , i=k+1,k+2,\cdots $.
They altogether span
the complete basis of ${\cal H}$.
We can label them by positive integer $n$ : 
\bea
\{ \, \, \left|\, h_n \, \right\ran , 
\, \,  n \in {\bf Z}_+  \} =
\{ \, \, \left|\, g_\a \, \right\ran , 
\, \left|\, f_i \, \right\ran , 
\, \a = 1,2, \cdots , k ,\, \, 
 i = k+1, k+2, \cdots  \, \, \}.
\ena
As we can see from (\ref{inclusion}),
zero-modes (\ref{zeroV})
are completely
determined by  
the ideal $f_i(z_1,z_2)$
\cite{LecNakaj}:
\bea
\left| {\cal U}(f_i) \right\ran
=
\left( 
\begin{array}{c}
  \left| u_1(f_i) \right\ran \\
  \left| u_2(f_i) \right\ran \\
  \left|\, f_i \, \right\ran 
 \end{array}
\right) .
\ena
We can construct 
operator zero-mode (\ref{zeroPsi})
by the following formula:
\bea
 \label{ozeroambi}
\Psi = 
\sum_{i}\sum_{n}
(\Psi)_{in}
\left| {\cal U}(f_i) \right\ran \left\lan h_n \right| ,
\ena
where
$(\Psi)_{in}$ is a 
commuting number.
From (\ref{ozeroambi}), one can see that
there are infinitely many operator
zero-modes.
Since 
the Fock space ${\cal H}$ is divided
into two orthogonal subspaces ${\cal H}_{\cal I}$ 
and ${\cal H}_{/{\cal I}}$ through the isomorphism (\ref{ideal}),
it is natural to restrict the action
of operators 
to ${\cal H}_{\cal I}$.
We call $\Psi_{0}$ ``minimal operator zero-mode" 
if it has the form:
\bea
& &\Psi_{0} = 
\sum_{i,j}
(\Psi_0)_{ij}
\left| {\cal U}(f_i) \right\ran 
\left\lan f_j \right| ,
\qquad \left| f_j \right\ran  \in {\cal H}_{\cal I}, \nn
& &\qquad \forall j, \, \,
\exists i \quad 
\mbox{such that} \quad
(\Psi_0)_{ij} \ne 0 ,
\ena
i.e.
$(\Psi_0 )_{ in} = 0$ for $n = \a = 1,2,\cdots ,k$.
We call it
``normalized" minimal operator zero-mode 
if it is normalized in ${\cal H}_{\cal I}$:
\bea
\Psi_{0}^{\dagger} \Psi_{0} 
= P_{\cal I}  ,
\ena
where $P_{\cal I}$ is a projection operator
which represents the projection to ${\cal H}_{\cal I}$,
the space of ideal states.
The uniqueness of the normalized minimal operator zero-mode 
(up to gauge transformation) is shown in appendix \ref{B}. 
This means that the normalized minimal 
operator zero-mode containes
minimal information of the ideal (\ref{ideal}).
From above definition, minimal operator zero-mode
annihilates states in ${\cal H}_{/{\cal I} }$, i.e.
$\Psi_0 \left| \varphi  \right\ran = 0 $ for
$\left| \varphi  \right\ran 
\in {\cal H}_{/{\cal I} }$.
Note that
if we write
\bea
 \label{xi0}
\Psi_0 =
\left(
\begin{array}{c}
  \psi_1 \\
  \psi_2 \\
  \xi
\end{array}
\right) ,
\ena
then
$\xi \left| \varphi  \right\ran =0 
\Rightarrow 
  \psi_1 \left| \varphi  \right\ran =
  \psi_2 \left| \varphi  \right\ran = 0
$.
Hence the states
annihilated by minimal operator zero-mode
$\Psi_0$ are completely determined
by the third factor $\xi $ in (\ref{xi0}).

An interesting point is that the
noncommutative operators
appear from the ideal described in purely
commutative terms,
by treating  infinite number of 
the elements of ideal simultaneously.


As an illustration, let us construct
$U(1)$ one-instanton solution from the ideal.
First, let us recall
$U(1)$ one-instanton solution 
constructed in \cite{NS}.
The solution to the modified ADHM equations
(\ref{ADHMzeta})
is given by
\bea
 \label{1-1}
B_1 = B_2 = 0 , 
\quad I = \sqrt{\zeta}\, , \, J = 0.
\ena
There is a solution $\tilde{\Psi}_0 $ to 
the equation 
${\cal D}_z \tilde{\Psi}_0 = 0 $ :
\bea
 \label{1-1zeroPsi}
\tilde{\Psi}_0 =
\left(
 \begin{array}{c}
  \tilde{\psi_1} \\
  \tilde{\psi_2} \\
  \tilde{\xi}
 \end{array}
\right) = 
\left(
 \begin{array}{c}
\sqrt{\zeta} \bar{z}_2 \\
\sqrt{\zeta} \bar{z}_1 \\ 
\left( z_1\bar{z}_1 + z_2 \bar{z}_2 \right) 
 \end{array}
\right)  .
\ena
Notice that all the components of
$\tilde{\Psi}_0 $ annihilate 
$\left| 0,0 \right\ran $ .
As a consequence 
$\tilde{\Psi}_0^{\dagger} \tilde{\Psi}_0 
= (z_1\bar{z}_1 + z_2\bar{z}_2 )
(z_1\bar{z}_1 + z_2\bar{z}_2 + \zeta)$
annihilates $\left| 0,0 \right\ran$.
Therefore the inverse of $(z_1\bar{z}_1 + z_2\bar{z}_2 )$
is only defined in the subspace 
of Fock space where
$\left| 0,0 \right\ran $ is projected out:
\EQ 
(z_1\bar{z}_1 + z_2\bar{z}_2 )^{-1}
:=
P
(z_1\bar{z}_1 + z_2\bar{z}_2 )^{-1} P,
\EN
where $P$ is a projection operator
that project out 
$\left| 0,0 \right\ran $:
\bea
P = \mbox{Id}_{\cal H} 
- \left|0,0 \right\ran \left\lan 0,0 \right|.
\ena
Therefore 
\bea
 \label{minzero}
\Psi_0
= \tilde{\Psi}_0 
(\tilde{\Psi}_0^{\dagger} 
\tilde{\Psi}_0 )^{-1/2}
\ena
is normalized as
$\Psi_0^{\dagger} \Psi_0 = P$.

Let us reconstruct this zero-mode
from ideal.
The ideal which corresponds to 
(\ref{1-1}) is
${\cal I}= ( z_1 , z_2) $.
The basis vector of 
the ${\cal H}_{/{\cal I}} $ is 
$\left| 0,0 \right\ran $ which is
orthogonal to 
all the ideal states.
We can use 
$\left| n_1, n_2   \right\ran , 
(n_1 , n_2 ) \ne (0,0) $
as 
basis vectors of ${\cal H}_{\cal I}$ ,
the space of ideal states.
The solutions of
${\cal D}_z \left| {\cal U}  \right\ran = 0 $
are given by
\bea
 \label{1-1V}
\left| {\cal U}_{n_1n_2} \right\ran  = 
\left(
 \begin{array}{c}
  \left| u_{1 n_1n_2} \right\ran \\
  \left| u_{2 n_1n_2} \right\ran \\
  \left|\, f_{n_1n_2} \,\right\ran 
 \end{array}
\right) =
\left(
 \begin{array}{c}
\sqrt{n_2}  \left| n_1 , n_2-1 \right\ran \\
\sqrt{n_1}  \left| n_1-1, n_2 \right\ran \\
\frac{1}{\sqrt{2} } (n_1 + n_2)
    \left| n_1, n_2   \right\ran 
 \end{array}
\right) , \quad (n_1, n_2 ) \ne (0,0) .
\ena
From (\ref{1-1V}), we obtain 
operator zero-mode 
${\cal D}_z \Psi = 0 $ :
\bea
 \label{1-1mPzero}
\Psi = 
\sum_{(m_1,m_2) \ne (0,0)}\sum_{(n_1,n_2)}
(\Psi)_{(m_1,m_2)(n_1,n_2)}
\left| {\cal U}_{m_1m_2} \right\ran 
\left\lan n_1,n_2 \right| ,
\ena
The normalized minimal operator zero-mode $\Psi_0 $
is required to satisfy
\bea
 \label{psizero}
& &\Psi_0 = 
\sum_{(m_1,m_2) \ne (0,0)}\sum_{(n_1,n_2)\ne (0,0)}
(\Psi)_{(m_1,m_2)(n_1,n_2)}
\left| {\cal U}_{m_1m_2} \right\ran 
\left\lan n_1,n_2 \right| , \nn
& &\Psi_0^{\dagger} \Psi_0 
=  \mbox{Id}_{\cal H} 
- \left| 0,0 \right\ran \left\lan 0,0 \right|  . 
\ena
From the normalization condition
in (\ref{psizero}),
we obtain
\bea
 \label{1-1m}
& &\sum_{(m_1,m_2) \ne (0,0)}
\frac{1}{2}(m_1 + m_2)(m_1 + m_2 + 2)\, 
(\Psi^{\dagger})_{(l_1,l_2)(m_1,m_2)}
(\Psi)_{(m_1,m_2)(n_1,n_2)}
=
\delta_{(l_1,l_2)(n_1,n_2)} . \nn
& & {}
\ena
The solution of (\ref{1-1m}) is
\bea
\label{matpsi}
(\Psi_0)_{ (m_1,m_2)(n_1,n_2)} = 
\sqrt{\frac{2}{(n_1 + n_2)(n_1 + n_2 + 2)} } 
\, \delta_{(m_1,m_2)(n_1,n_2)}\, \, . 
\ena
(\ref{psizero}) and
(\ref{matpsi}) are equivalent to 
(\ref{1-1zeroPsi}) and (\ref{minzero}).


\subsection{Some $U(1)$ Instanton Solutions}


The construction of operator zero-mode from
vector zero-modes is useful 
for understanding
the notion of minimal operator zero-mode.
But in some simple cases
it is 
easier to directly look for the 
operator zero-modes.
It is interesting to observe that
the obtained operator zero-modes which are
most naturally obtained
really annihilate states 
in ${\cal H}_{/{\cal I} }$.

\subsection*{$U(1)$ 
two-instanton 
solution\footnote{Although we only
consider
solutions of matrix equation
(\ref{ADHMzeta}) and
do not construct
gauge field explicitly,
we call the solutions
``instanton solutions"
because in principle
we can construct instantons
from the matrix data.
We regard $k$ as a number of
instantons.
} }

Let us study the two-instanton solutions
degenerating at the origin.
The corresponding solution to the
matrix equations (\ref{ADHMzeta}) is 
given by
\bea
 \label{1-2}
B_1 =
\left(
\begin{array}{cc}
 0 & \sqrt{\zeta} \lambda_1 \\
 0 & 0 
\end{array}
\right) , \, \,
B_2 =
\left(
\begin{array}{cc}
 0 & \sqrt{\zeta} \lambda_2 \\
 0 & 0 
\end{array}
\right) , \, \,
I = 
\left(
\begin{array}{c}
 0 \\
 \sqrt{2\zeta}
 \end{array} 
\right) 
, \, \, J = 0 . 
\ena
where $\lambda_1 $ and $\lambda_2$ are
complex numbers satisfying
$|\lambda_1|^2 + |\lambda_2|^2 = 1 $ .
Notice that $B_1$ and 
$B_2$ are upper half triangle matrices.
$\lambda_1 , \lambda_2 $ 
(partially) remember
the direction before two instantons
collide 
\cite{LecNakaj}\cite{Ours}.
The corresponding ideal is
$
{\cal I} = 
\bigl( z_1^2 , \, 
- \lambda_2 z_1 + \lambda_1 z_2 \bigr)
$.
Hence the states 
orthogonal to all the ideal states
are annihilated by
$
\bar{z}_1^2 , \, \, 
-\lambda_2^* \bar{z}_1 
+ \lambda_1^* \bar{z}_2 .
$
The states annihilated by
$\bar{z}_1^2 $ is
$
\left| 0, n_2  \right\ran , \, 
\left| 1, n_2  \right\ran 
$
for all non-negative integer $n_2$.
In order to describe the
states annihilated by
$-\lambda_2^* \bar{z}_1 
+\lambda_1^*  \bar{z}_2 $,
it is simpler to use the basis
constructed by rotated creation and annihilation
operators $z'$ and $\bar{z}'$:
\bea
\label{rot}
& &z_1' \equiv 
\lambda_1^*  z_1 
+ \lambda_2^*  z_2 , \qquad
z_2' \equiv
-\lambda_2 z_1 
+ \lambda_1  z_2  ,          \nn
\vs{5}
& &  \left| 0, 0 \right\ran
= \left| 0, 0 \right\ran_{\lambda} ,              \nn
& &\sqrt{\frac{2}{\zeta} } z_1' 
\left| n_1'  , n_2'    \right\ran_{\lambda} 
= \sqrt{n_1' +1} 
\left| n_1' + 1 , n_2'    \right\ran_{\lambda}  , \quad
\sqrt{\frac{2}{\zeta} } \bar{z}_1'
\left| n_1'  , n_2'    \right\ran_{\lambda} 
= \sqrt{n_1' } \left| n_1' -1 , n_2'  
\right\ran_{\lambda} ,          \nn
& &\sqrt{\frac{2}{\zeta} } z_2' 
\left| n_1'  , n_2'    \right\ran_{\lambda} 
= \sqrt{n_2' +1} \left| n_1' , n_2' +1    
\right\ran_{\lambda} , \quad
\sqrt{\frac{2}{\zeta} }
\bar{z}_2' \left| n_1'  , n_2'    \right\ran_{\lambda} 
= \sqrt{n_2'} \left| n_1' , n_2' -1    
\right\ran_{\lambda} . \nn
& & {}
\ena
Then the states annihilated by
$\bar{z}_2' = -\lambda_2^* \bar{z}_1 
+ \lambda_1^*  \bar{z}_2 $ are
$
 \left|  n_1' , 0 \right\ran_{\lambda}
$
for all non-negative
$n_1'$ .
Therefore the basis 
vectors
of the 
states orthogonal
to all the ideal states are
$
\left| 0 , 0 \right\ran , 
\left| 1 , 0 \right\ran_{\lambda}
$
Now let us study operator zero-mode.
The (unnormalized) 
minimal operator zero-mode 
can be directly
obtained from (\ref{1-2}):
\bea
 \label{1-2zero}
& &
\tilde{\Psi}_0 =
\left(
\begin{array}{c}
\tilde{\psi_1} \\
\tilde{\psi_2} \\
\tilde{\xi}
\end{array}
\right), \qquad 
\tilde{\psi}_1 =
\left(
\begin{array}{c}
\sqrt{\zeta}\bar{z}_2 \bar{z}_1'  \\
\bar{z}_2  
\frac{\zeta}{2}( \hat{N} - 1 ) 
+ \zeta \lambda_1 
\bar{z}_2' 
\end{array}
\right)\,  ,\nn
& & \qquad \qquad \qquad \quad \quad
\tilde{\psi}_2
=
\left(
\begin{array}{c}
\sqrt{\zeta}
\bar{z}_1 
\bar{z}_1'  \\
\bar{z}_1
\frac{\zeta}{2} (\hat{N}-1)
- \zeta
\lambda_2 
\bar{z}_2'
\end{array}
\right)\, , \nn
& &
\qquad \qquad \qquad \quad \quad
\tilde{\xi} =
\frac{1}{\sqrt{2 \zeta } }
\left( \frac{\zeta}{2} \right)^2
\left(
\hat{N}(\hat{N} - 1 ) + 2 n_2'
\right) \,  ,
\ena
where 
$ \frac{\zeta}{2}\hat{N} 
   \equiv z_1 \bar{z}_1 + z_2 \bar{z}_2 $, 
$ \frac{\zeta}{2}\hat{n}_2' \equiv
    z_2' \bar{z}_2' $.
(\ref{1-2zero}) is really minimal:
$ \left| 0,0 \right\ran $ and 
$\left| 1 , 0 \right\ran_{\lambda}$ 
are annihilated 
by all the components of 
$\tilde{\Psi}_0 $.

\subsection*{$U(1)$
three-instanton solutions}

Let us consider the $k=3$ solution
corresponding to the following simple ideal: 
\footnote{This kind of ideal corresponds to a 
fixed points of $T^2$ action 
in \cite{Nakaj}\cite{LecNakaj}.  }
\bea
{\cal I} =
\left\{ f(z_1,z_2) = \sum_{n_1,n_2}
a_{n_1n_2} z_1^{n_1} z_2^{n_2} \Biggm|
\begin{array}{l}
a_{n_1n_2} = 0\,  
\mbox{ when $(n_1,n_2)$ belongs } \\
\mbox{to the Young tableau (Y$1$).  }
\end{array}
\right\}        
\ena
\begin{equation}
 \begin{array}{ccc} 
 \cline{1-1}
  \svline{(1,0)}&\svline{} &\\
 \cline{1-2}
 \svline{(0,0)} & \svline{(0,1)} &\svline{}   \\ 
 \cline{1-2}
  &(\mbox{Y}1) & 
 \end{array}
\end{equation}
The solution to (\ref{ADHMzeta}) is given by
\bea
 \label{1-3mat1}
B_1 =
\left(
\begin{array}{ccc}
 0 & 0 & 0 \\
 0 & 0 & \sqrt{\zeta} \\
 0 & 0 & 0 
\end{array}
\right), \quad
B_2 =
\left(
\begin{array}{ccc}
 0 & 0 & \sqrt{\zeta} \\
 0 & 0 & 0 \\
 0 & 0 & 0 
\end{array}
\right) , \quad
I  =
\left(
\begin{array}{c}
 0  \\
 0  \\
 \sqrt{ 3 \zeta} 
\end{array}
\right) , \quad J = 0 .
\ena
We can find the 
(unnormalized)
minimal operator zero-mode: 
\bea
 \label{3zero1}
& &
\tilde{\Psi}_0 =
\left(
\begin{array}{c}
 \tilde{\psi}_1 \\
 \tilde{\psi}_2 \\
  \tilde{\xi}
\end{array}
\right), \quad
\tilde{\psi}_1 =
\left(
\begin{array}{c}
 \sqrt{\zeta} \bar{z}_2^2 \\
 \sqrt{\zeta} \bar{z}_1\bar{z}_2\\
 \frac{\zeta}{2} \hat{N} \bar{z}_2
\end{array}
\right)   , \quad
\tilde{\psi}_2 =
\left(
\begin{array}{c}
 \sqrt{\zeta} \bar{z}_1\bar{z}_2 \\
 \sqrt{\zeta} \bar{z}_1^2 \\
 \frac{\zeta}{2} \hat{N} \bar{z}_1
\end{array}
\right)  ,  \nn
& & \qquad \qquad \qquad \quad
\tilde{\xi} =
\frac{1}{\sqrt{3 \zeta} }
\left( 
\frac{\zeta}{2}
\right)^2
\hat{N}(\hat{N} - 1)  .
\ena
(\ref{3zero1}) really annihilates
$
\left| 0,0 \right\ran , \,
\left| 1,0 \right\ran , \,
\left| 0,1 \right\ran 
$
and hence minimal.

Next consider the ideal
corresponding to the following 
Young tableau (Y$2$):
\begin{equation}
 \begin{array}{cc} 
 \cline{1-1}
  \svline{(2,0)}&\svline{}  \\
 \cline{1-1}
  \svline{(1,0)}&\svline{}   \\
 \cline{1-1}
 \svline{(0,0)} &\svline{}   \\ 
 \cline{1-1}
 (\mbox{Y}2) &
 \end{array}
\end{equation}
The solution to (\ref{ADHMzeta}) is
given by
\bea
B_1 =
\left(
\begin{array}{ccc}
 0 & \sqrt{\zeta} & 0 \\
 0 & 0 & \sqrt{2\zeta}  \\
 0 & 0 & 0
\end{array}
\right) , \, \, 
B_2 = 0 , \, \,
I =
\left(
\begin{array}{c}
  0 \\
  0 \\
\sqrt{3 \zeta}
\end{array}
\right) , \, \, J = 0 .
\ena
The (unnormalized) minimal operator 
zero-mode is given as
\bea
\label{3zero2}
& &
\tilde{\Psi}_0 =
\left(
\begin{array}{c}
 \tilde{\psi}_1 \\
 \tilde{\psi}_2 \\
  \tilde{\xi}
\end{array}
\right), \quad
\tilde{\psi}_1 =
\left(
\begin{array}{c}
 2\zeta  \bar{z}_1^2 \bar{z}_2 \\
 \sqrt{2\zeta} 
 \frac{\zeta}{2} \hat{N} 
 \bar{z}_1\bar{z}_2\\
\left(\frac{\zeta}{2}\right)^2
\left\{ (\hat{N}+ 1)(\hat{N}+ 4) 
- 2 (\hat{n}_1 - 1)  
\right\}
\bar{z}_2
\end{array}
\right)   , \nn
& &\qquad \qquad \quad \quad
\tilde{\psi}_2 =
\left(
\begin{array}{c}
 2 \zeta \bar{z}_1 ^3 \\
 \sqrt{2\zeta} 
 \frac{\zeta}{2} \hat{N} \bar{z}_1 ^2   \\
\left( \frac{\zeta}{2} \right)^2
\left\{ (\hat{N}+ 1)\hat{N} 
- 2 \hat{n}_1  
\right\}
\bar{z}_1
\end{array}
\right)  ,  \nn
& & \qquad \qquad \qquad \quad
\tilde{\xi} =
\frac{1}{\sqrt{3 \zeta} }
\left( 
\frac{\zeta}{2}
\right)^3
\hat{N}
\left\{
\hat{N}  (\hat{N} +3 ) 
- 2 (3 \hat{n}_1 - 1 )
\right\} .
\ena
We can check
(\ref{3zero2}) annihilates
$
\left| 0,0 \right\ran , \,
\left| 1,0 \right\ran , \,
\left| 2,0 \right\ran .
$

\section{$U(N)$ Instantons
and Projection Operators}\label{Un}

In the previous section we
have clarified the
notion of the minimal operator zero-mode
for $U(1)$ case.
In this section
we will study the 
$U(2)$ instanton solutions
and observe that the projection
of states by zero-modes also occurs. 
Since the $U(N)$ instanton solutions
are essentially embeddings of $U(2)$
instanton solutions to $U(N)$,
this means that the projection
of states is a general phenomenon
in the ADHM construction of
instantons on noncommutative
${\bf R}^4$.
In the following, we will make two
observations:
\begin{enumerate}
 \item The minimal operator zero-mode appears
in the $U(1)$ subgroup of $U(2)$ gauge group.
It annihilates some states even when the size of instanton
is not small.
 \item When the size of instanton becomes
small, 
only the contribution from
$U(1)$ subgroup 
described by the
minimal operator zero-mode remains.
\end{enumerate}
Although we have not 
defined minimal operator zero-mode
for $U(N)$ case, zero-modes
similar to the minimal operator zero-mode
in $U(1)$ case appear
in explicit solutions.
Hence in the above we have also 
called them minimal operator zero-modes.
The second observation can be 
understood as follows.
We may define
``small instanton" on noncommutative
${\bf R}^4$ as 
$J = 0$ solution of the 
modified ADHM equations 
(\ref{ADHMzeta}).
Then the solution is
essentially the 
embedding of $U(1)$ 
instanton to $U(N)$.

\subsection*{$U(2)$ one-instanton solution}

The solution to the modified
ADHM equations (\ref{ADHMzeta}) is given 
by\footnote{%
There are of course
family of solutions
with different orientation
in gauge group $U(2)$. 
The resulting conclusions are the same.
}
\bea
B_1 = B_2 = 0 ,
\quad
& &I = \left(
\begin{array}{cc}
\sqrt{\rho^2 + \zeta} &
 0
\end{array}
\right)       , \quad
J^{\dagger }
=
 \left(
\begin{array}{cc}
      0 &
  \rho 
\end{array}
\right) ,
\ena
where 
$\rho$ 
is a real non-negative number
and parameterizes the 
size of the instanton.
The two ``orthonormalized"
operator zero-modes of ${\cal D}_z $
are given by
\bea
& &\Psi^{(1)}=
\left(
\begin{array}{c}
\psi_1^{(1)} \\
\psi_2^{(1)} \\
\xi^{(1)} \\
{}
\end{array}
\right)
=\left(
\begin{array}{c}
\sqrt{\rho^2 + \zeta}\, \bar{z}_2 \\
\sqrt{\rho^2 + \zeta}\, \bar{z}_1\\
(z_1 \bar{z}_1 + z_2 \bar{z}_2)  \\
    0
\end{array}
\right)
\left(
(z_1\bar{z}_1 + z_2 \bar{z}_2 )
(z_1\bar{z}_1 + z_2 \bar{z}_2 
  + \zeta + \rho^2)
\right)^{-1/2}
 ,     \label{2-1zero1}  \\
& &\Psi^{(2)}
=
\left(
\begin{array}{c}
\psi_1^{(2)} \\
\psi_2^{(2)} \\
\xi^{(2)} \\
{}
\end{array}
\right)
=\left(
\begin{array}{c}
- \rho z_1 \\
\rho z_2 \\
   0 \\
( z_1\bar{z}_1 + z_2\bar{z}_2 
+ \zeta ) 
\end{array}
\right) 
\left(
\left( z_1\bar{z}_1 +
z_2\bar{z}_2  + \zeta \right)
\left( z_1\bar{z}_1 +
 z_2\bar{z}_2  
+  \zeta +\rho^2 \right) 
\right)^{-1/2}. \nn
& & {} \label{2-1zero2}
\ena
The zero-mode $\Psi^{(1)}$
is a straightforward modification
of (\ref{minzero}).
$\Psi^{(1)}$ annihilates
$\left|0,0 \right\ran$ for
any values of $\rho$,
and normalized in the subspace
where
$\left|0,0 \right\ran$ is 
projected out.
The zero-mode $\Psi^{(2)}$
annihilates no state in
${\cal H}$ and 
manifestly non-singular
even if $\rho = 0$.
When 
$\rho = 0$,
$\psi_1^{(2)} =  \psi_2^{(2)} = 0$ ,
and from (\ref{FS}) 
$\Psi^{(2)}$ does not contribute
to the field strength. 
Therefore
the structure of the
instanton at $\rho = 0$
is completely determined
by the
$U(1)$ subgroup described
by minimal operator zero-mode $\Psi^{(1)}$.

\subsection*{$U(2)$ two-instanton solution}

We can also construct a two-instanton solution and 
check the statements in the beginning of this section.
Here we only construct one simple
solution.
The solution of 
modified ADHM equations (\ref{ADHMzeta})
is given by
\bea
B_1 =
\left(
\begin{array}{cc}
 0 & \sqrt{\zeta} \\
 0 & 0
\end{array}
\right) , \, \, B_2 = 0, \quad
I =
\left(
\begin{array}{cc}
 0 & 0 \\
 \sqrt{2(\rho^2 + \zeta) } & 0
\end{array}
\right) , \, \,
J^{\dagger }=
\left(
\begin{array}{cc}
 0 & 0 \\
 0 & \rho
\end{array}
\right) .
\ena
We can obtain two (unnormalized)
zero-modes orthogonal to each other:
\bea
& &\Psi^{(1)}
=
\left(
\begin{array}{c}
\psi_1^{(1)} \\
\psi_2^{(1)} \\
 \xi^{(1)}
\end{array}
\right),   \qquad
\psi_1^{(1)}
 =
\left(
\begin{array}{c}
\sqrt{\rho^2 + \zeta}
\sqrt{\zeta} \bar{z}_1\bar{z}_2 \\
\sqrt{\rho^2 + \zeta}
\bar{z}_2 
(z_1\bar{z}_1 + z_2\bar{z}_2 
+ \frac{\zeta}{2} )
\end{array}
\right)   , \nn
& & \qquad \qquad \qquad \qquad \qquad
\psi_2^{(1)}
 =
\left(
\begin{array}{c}
\sqrt{\rho^2 + \zeta}
\sqrt{\zeta} \bar{z}_1^2 \\
\sqrt{\rho^2 + \zeta}
\bar{z}_1
(z_1\bar{z}_1 + z_2\bar{z}_2 
- \frac{\zeta}{2} ) \\
\end{array}
\right) , \nn
& & \qquad \qquad \qquad \quad
\xi^{(1)} =
\left(
\begin{array}{c}
\frac{1}{\sqrt{2}}
(z_1\bar{z}_1 + z_2\bar{z}_2 )
(z_1\bar{z}_1 + z_2\bar{z}_2 
- \frac{\zeta}{2})
+ \zeta z_2 \bar{z}_2 \\
 0 
\end{array}
\right),
\ena
and
\bea
& &\Psi^{(2)}
=
\left(
\begin{array}{c}
\psi_1^{(2)} \\
\psi_2^{(2)} \\
 \xi^{(2)}
\end{array}
\right),   \qquad
\psi_1^{(2)}
 =
\left(
\begin{array}{c}
\rho
\sqrt{\zeta} \bar{z}_2 z_2 \\
\rho
z_1 
(z_1\bar{z}_1 + z_2\bar{z}_2 
+ \frac{\zeta}{2} )
\end{array}
\right)   , \nn
& & \qquad \qquad \qquad \qquad \qquad
\psi_2^{(2)}
 =
\left(
\begin{array}{c}
\rho 
\sqrt{\zeta} \bar{z}_1 z_2 \\
\rho
z_2
(z_1\bar{z}_1 + z_2\bar{z}_2 
+ \frac{\zeta}{2} ) \\
\end{array}
\right) , \nn
& & \qquad \qquad \quad 
\xi^{(2)} =
\left(
\begin{array}{c}
 0 \\
\frac{1}{\sqrt{2}}
\left(
(z_1\bar{z}_1 + z_2\bar{z}_2 )
(z_1\bar{z}_1 + z_2\bar{z}_2 
+ \frac{\zeta}{2})
+ \zeta (z_2 \bar{z}_2 +\frac{\zeta}{2})
\right)
\end{array}
\right) .
\ena
$\Psi^{(1)} $ is a slight
modification of (\ref{1-2zero})
with $(\lambda_1,\lambda_2) = (1,0)$.
It annihilates
$\left| 0,0
 \right\ran ,
\left| 1,0
 \right\ran $.
$\Psi^{(2)} $ is 
apparently
non-singular and 
$\psi^{(2)}_1 =\psi^{(2)}_2 =0$ when $\rho =0$.
Hence when the size
of the instanton is small,
only the $U(1)$ subgroup described
by $\Psi^{(1)} $ contributes 
to the field strength.


\section{D-Instanton Makes 
a Hole on D3-Brane}\label{secIIB}

The existence of the projection
operator forces us
to consider the reduced Fock space.
In this section it is shown that
the projection can be interpreted as
modification of spacetime topology.
Usual Yang-Mills theory cannot
describe such
spacetime topology change.
However, as we will see shortly, 
IIB matrix model 
\cite{IKKT}\cite{IKKTB} gives
an appropriate framework.
The action of the
IIB matrix model is obtained by
dimensionally reducing 
ten-dimensional $U(N)$ 
super Yang-Mills theory
down to zero dimension:\footnote{%
We have slightly changed the notations from 
those in the previous sections:
in this section $N$ denotes 
the rank of the gauge group of IIB matrix model.
We will only consider 
$U(1)$ instantons in the following.
}
\bea
 \label{IIB}
S = -\frac{1}{g^2}
\mbox{Tr} 
\left(
\frac{1}{4}[X_\mu, X_\nu][X^\mu, X^\nu]
+
\frac{1}{2}\bar{\Theta}\Gamma_\mu
[X^\mu,\Theta]
\right),
\ena
where 
$X_\mu$ and $\Theta$ are $N \times N$
hermitian matrices and each component
of $\Theta$ is a Majorana-Weyl spinor.
The action (\ref{IIB}) 
has the following ${\cal N}=2$ supersymmetry:
\bea
\delta^{(1)} \Theta 
&=& \frac{i}{2} [X_{\mu},X_{\nu}] 
\Gamma^{\mu\nu} \epsilon^{(1)}, \nn  
\delta^{(1)} X_\mu 
&=& i\bar{\epsilon}^{(1)} \Gamma_{\mu} \Theta , \nn
\delta^{(2)} \Theta &=& \epsilon^{(2)}, \nn
\delta^{(2)} X_\mu &=& 0.
\ena
The classical equation of motion is given by
\bea
 \label{IIBeq}
[X_{\mu},[X_{\mu},X_{\nu}]] =0.
\ena 
IIB matrix model has
classical
D-brane solutions:
\bea
 \label{half}
& &X_\mu = i \hat{\pa}_\mu,\nn
& &[i\hat{\pa}_\mu, i\hat{\pa}_\nu] = - i B_{\mu\nu},
\ena
where $B_{\mu\nu}$'s are real constants.
Hereafter we will consider
(Euclidean) D3-brane solution, 
i.e.
the rank of $B_{\mu\nu}$ is four
and $B_{\mu\nu}= 0$ when 
$\mu,\nu \ne 1,2,3,4$.
We define ``coordinate matrices" $\hat{x}^{\mu}$ by
\bea
\hat{x}^{\mu} = - i \theta^{\mu \nu}\hat{\pa}_{\nu},
\ena
where $\theta^{\mu\nu}$ is an inverse
matrix of $B_{\mu\nu}$.
Then their commutation relations are the   
same as those in (\ref{noncomx}): 
\bea
[\hat{x}^{\mu}, \hat{x}^{\nu} ] = i \theta^{\mu\nu}.
\ena
Hence by setting $\theta^{\mu\nu}$
(or equivalently  $B_{\mu\nu}$) self-dual as in 
(\ref{theta}), i.e.
$\theta^{12} = \theta^{34} 
= \frac{\zeta}{4}$, and replacing
operators to 
infinite rank 
matrices,\footnote{%
(\ref{half}) is not satisfied in $U(N)$ IIB matrix
model with finite $N$.
}
we can embed 
the instanton solution (\ref{PAP})
to IIB matrix model:
\bea
 \label{four}
X_{\mu} 
= P(i\hat{\pa}_{\mu} + i A_{\mu})P
\ena
where $A_\mu$ is the $U(1)$
instanton solution obtained
by ADHM construction:
\bea
A_{\mu} 
= \Psi^{\dagger}[\hat{\pa}_{\mu},\Psi]P ,
\ena
where $\Psi$ is a zero-mode (\ref{zeroPsi}).
$P$ is the projection operator 
determined by the zero-mode,
as described in section \ref{secIdeal}.
From (\ref{four})
the solution can be represented
within reduced Fock space
$P{\cal H}: 
= \sum_{(n_1,n_2) \in {\bf Z}^2_{\geq 0} }
{\bf C} (P \left|  n_1,n_2 \right\ran ) $.
Therefore the solution is realized 
by $N \times N$ matrices
with $N = (\mbox{dim}{\cal H}- k )$,
where $k$ is an instanton number.
Notice that in (\ref{four})
instanton and geometry(D3-brane) 
are combined into
single solution. Indeed, we can rewrite
(\ref{four}) into simpler form:
\bea
 \label{IIsimple}
X_{\mu} 
&=& P(i\hat{\pa}_{\mu} + i A_{\mu})P \nn
&=& P(i\hat{\pa}_{\mu})P 
+ P(i \Psi^{\dagger} \hat{\pa}_\mu \Psi )P -
P(i \Psi^{\dagger} \Psi P \hat{\pa}_{\mu})P \nn
&=& i P \Psi^{\dagger} \hat{\pa}_\mu \Psi P 
= i \Psi^{\dagger} \hat{\pa}_\mu \Psi  .
\ena
From (\ref{four}) we obtain
\bea
 \label{XX}
[X_{\mu},X_{\nu}]
=
P(-i B_{\mu\nu} - F^-_{\mu\nu \mbox{\tiny ADHM}})P.
\ena
The derivation is
similar to (\ref{PAP}) $\sim$ (\ref{FP})
and $F^-_{\mu\nu \mbox{\tiny ADHM}}$ is anti-self-dual.
From (\ref{XX}) it is easy to check
that $X_\mu$ in (\ref{IIsimple})
solves the 
equation of motion (\ref{IIBeq}).

Let us consider the
supersymmetry transformation
in this background:
\bea
 \label{SUSYsol}
\delta^{(1)} \Theta 
&=&
\frac{i}{2}[X_\mu,X_\nu] \Gamma^{\mu\nu} \epsilon^{(1)} \nn
&=&
\frac{i}{2}
P(-i B_{\mu\nu}
-F^-_{\mu\nu \mbox{\tiny ADHM}}\frac{1+\Gamma_5}{2})P 
\Gamma^{\mu\nu}\epsilon^{(1)}, \nn
\delta^{(2)} \Theta &=& \epsilon^{(2)}.
\ena
From (\ref{SUSYsol}) we 
can see that the solution (\ref{four}) 
preserves one fourth of supersymmetry \cite{IKKTB}:
\bea
 \label{SUSYsol2}
\Gamma_5 \epsilon^{(1)} &=& -  \epsilon^{(1)} ,  \nn
\epsilon^{(2)} 
&=& 
- \frac{1}{2}
P B_{\mu\nu}P \Gamma^{\mu\nu}
\epsilon^{(1)}.  
\ena
Notice that the projection operator
is an identity operator in the 
reduced Fock space $P{\cal H}$.
Hence the second
supersymmetry transformation
is proportional to the
identity matrix in $U(N)$ IIB matrix model,
with $N = \mbox{dim} {\cal H} - k$.

The physical interpretation 
of the projection 
in this setting
is as follows.
The $B_{\mu\nu}$ in (\ref{half})
is interpreted as NS-NS B-field on
D3-brane worldvolume \cite{IKKTB}.
We have set $B_{\mu\nu}$
{\em self-dual}.
Since the self-dual
$B$-field on D3-brane
induces
negative D-instanton charge,\footnote{%
Our convention is:
D-instanton $\sim$ instanton $\sim$ 
{\em anti}-self-dual
.}
we can regard that the
D3-brane is made 
of infinitely many
constituent 
anti-D-instantons.
Now let us consider 
D-instantons within these
infinite number of anti-D-instantons.
In order for this configuration to
become BPS, 
it is necessary to change the configurations of
constituent anti-D-instantons.
The projection removes 
anti-D-instantons at the place 
of D-instantons and makes holes on
D3-brane worldvolume.

We can express
the holes made by
the projections by
rewriting above operator formulas 
using ordinary functions and star-product.
More precisely, we map
operators to normal symbols (see appendix \ref{Nsymbol}).
\footnote{%
Here we use normal symbols only
to give concrete expressions of holes
on ${\bf R}^4 \approx {\bf C}^2$.
It may be interesting to 
formulate field theory on noncommutative
${\bf R}^4$ using normal symbols.
It may be also interesting
to investigate the 
relation to superstring theory.
These are, however, beyond the scope of this paper
.}
For example, consider the projection 
corresponding to the ideal
${\cal I}$ generated by 
$(z_1-w_1^i ,z_2-w_2^i  )  \, (i = 1, \cdots, k)$.
Then, all normal symbols corresponding to
operators acting in the reduced Fock space
$\mbox{End} P{\cal H} $
vanish at $ (z_1,z_2) =(w_1^i, w_2^i)
\, (i = 1, \cdots, k)$. 
This can be equivalently
stated as the points
$ (z_1,z_2) =(w_1^i, w_2^i)\, (i = 1, \cdots, k)$
do not exist, or appear as holes. 

Using operator symbols,
one can show that the projection
removes $k$ units of anti-D-instanton charge. 
Let us calculate
(anti-)instanton number when there are no
D-instantons.
Using 
(\ref{intr}),
\bea
& &\frac{1}{16\pi^2}
\int d^4x
B_{\mu\nu} \tilde{B}^{\mu\nu} 
=
\frac{1}{16\pi^2}
\left( 2\pi \frac{\zeta}{4} \right)^2
\mbox{Tr}_{\cal H}\, 
4 \left( \frac{4}{\zeta} \right)^2
=
\mbox{Tr}_{\cal H}, \nn
& &\qquad \tilde{B}_{\mu\nu}
= \frac{1}{2}
{\epsilon_{\mu\nu}}^{\rho\sigma}B_{\rho\sigma}.
\ena
Since the
projection reduces dimension of Fock space
by $k$, it reduces $k$
units of anti-D-instanton charge
(Of course 
there are also contributions from anti-self-dual part.
Here we only mention the role of the projection).
This fact also supports
the idea that
the projection removes anti-D-instantons.

\section*{Conclusions and
Speculations}

\subsection*{Conclusions}
In this paper we have learned
that the appearance of 
projection operators
is a general phenomenon
in the ADHM construction 
on noncommutative
${\bf R}^4$.
It has been shown how to
treat these projections.
The existence of
the projection operator
forces us to consider gauge fields
on reduced Fock space.
Since noncommutative
${\bf R}^4$ is defined by algebra 
over whole Fock space, the projection means
the change of the spacetime
topology from (noncommutative) ${\bf R}^4$.
In order to
describe such change of
spacetime topology,
it seems appropriate
to consider theory which can describe 
both gauge theory and geometry.
Therefore we have embeded
the instanton solution to
IIB matrix model.
In IIB matrix model, 
instanton and geometry are
combined into single classical solution.

\subsection*{Speculations}
In \cite{BN} it was conjectured
that the $U(1)$ instanton on noncommutative
${\bf R}^4 \approx {\bf C}^2$ can be transformed 
to $U(1)$ instanton on
commutative K\"ahler manifold
which is a blowup of ${\bf C}^2$,
via field redefinition described in \cite{SW}.
The ideal used to describe 
projection in this paper is essentially the same
as the one used to describe
blowup in \cite{BN}.
Since both instantons are constructed
from the same ADHM data, the correspondence
is of course one-to-one.
It is interesting if this
correspondence is understood as 
field redefinition along the lines of \cite{SW}.

In section \ref{secIIB}
we have embeded instanton solutions
to IIB matrix model.
Instantons on noncommutative
${\bf R}^4$ represent
D-instantons within D3-brane worldvolume.
We interpret D3-brane as
bound states of infinitely
many anti-D-instantons.
Then the bound states of
D-instantons and D3-brane
are interpreted as
bound states of
D-instantons and anti-D-instantons.
As shown in (\ref{SUSYsol}) and (\ref{SUSYsol2}),
this co-existence of positive and negative
D-instanton charges still preserves
one fourth of supersymmetry.
However, anti-D-instantons are removed
at the place of D-instantons.
This fact strongly
suggests the relation
to brane-anti-brane
pair annihilation \cite{Sen}.
IIB matrix model describe above
D-instanton-D3-brane bound states
simply as its classical solution.
This fact indicates the power of 
IIB matrix model in the description 
of the fate of brane-anti-brane
unstable systems.
It is also straightforward to embed the 
noncommutative instanton solution
to BFSS matrix model. 
It is interesting to
study the instanton solution 
in IIB matrix model or BFSS matrix model
from the point of view of
brane-anti-brane pair annihilation \cite{AH2}.
In order to classify the
topological charges which 
will be preserved during pair annihilations,
investigations from K-theoretical viewpoints
may be important \cite{K1}\cite{K2}.

From above considerations,
D3-brane may be regarded as a kind of
``Dirac sea" for D-instantons.
This gives new viewpoints to the 
second quantization of branes 
\cite{Ours}\cite{nakatsu}.

\section*{Acknowledgments}

I am very grateful to T. Nakatsu for valuable discussions,
especially on the relation to the ideal.
I would also like to thank T. Kubota for reading 
the manuscript.

\newpage

\appendix


\section{The Absence of Zero-mode
of $\Box_z $ }\label{A}

In this section we show
that $\Box_z $ in (\ref{box}) has
no zero-mode.
Suppose
\bea
\Box_z  \left| v \right\ran = 0
\ena
for some
$ \left| v \right\ran $ ,
where
$ \left| v \right\ran 
\in {\cal H}^{\oplus k}$ , i.e.
$ \left| v \right\ran $ is vector
in $V = {\bf C}^k $
and vector in ${\cal H}$.
Then,
\bea
& &\left\lan v \right| \Box_z \left| v \right\ran = 0 \nn
&\Rightarrow &
\, \left\lan v \right| 
\tau_z
\tau_z^{\dagger}
\left| v \right\ran  \nn
& &=
\left\lan v \right| 
( B_1-z_1 )
( B_1^{\dagger}-\bar{z}_1 )
\left| v \right\ran 
+
\left\lan v \right|
( B_2-z_2 )
( B_2^{\dagger}-\bar{z}_2 )
\left| v \right\ran 
+
\left\lan v \right|
I I^{\dagger}
\left| v \right\ran 
= 0 , \nn
& & \, \left\lan v \right| 
\sigma_z^{\dagger } 
\sigma_z
\left| v \right\ran \nn
& &= 
\left\lan v \right|
( B_1^{\dagger}-\bar{z}_1 )
( B_1-z_1 )
\left| v \right\ran 
+
\left\lan v \right|
( B_2^{\dagger}-\bar{z}_2 )
( B_2-z_2 )
\left| v \right\ran 
+ 
\left\lan v \right|
J^{\dagger } J 
\left| v \right\ran = 0   .\nn
\ena
Since the norm of vectors in $V$ are
non-negative,
\bea
 \label{nonnega}
& &( B_1^{\dagger}-\bar{z}_1 )
\left| v \right\ran = 0 , \quad
( B_2^{\dagger}-\bar{z}_2 )
\left| v \right\ran = 0 , \quad
I^{\dagger} \left| v \right\ran  = 0 ,  \nn
& &( B_1-z_1 )
\left| v \right\ran =0 , \quad
( B_2-z_2 )
\left| v \right\ran =0 , \quad
J \left| v \right\ran = 0 . 
\ena
From (\ref{nonnega}), we obtain
\bea
\left\lan v \right| 
\zeta
\left| v \right\ran 
&=&
\left\lan v \right| 
[ B_1 , B_1^{\dagger} ] + [ B_2 , B_2^{\dagger} ]
+ II^{\dagger} - J^{\dagger} J
\left| v \right\ran  \nn
&=&
\left\lan v \right| 
[z_1, \bar{z}_1] + [z_2, \bar{z}_2]
\left| v \right\ran \nn
&=&
-
\left\lan v \right| 
\zeta
\left| v \right\ran.
\ena
This means
$\left| v \right\ran  =0 $.

\section{The 
Uniqueness of the
Normalized Minimal \\ Operator Zero-mode}\label{B}

In this appendix we show the
uniqueness of the normalized
minimal operator zero-mode 
(up to gauge 
transformation)
when the gauge group is $U(1)$.
Let us consider operator
zero-mode which has 
the following form:
\bea
\Psi_0
= \sum_{i,j}
(\Psi_0)_{ij}
\left| {\cal U}(f_i) \right\ran
\left\lan  f_j   \right| .
\ena
Then its norm is:
\bea
\Psi_0^{\dagger} \Psi_0
=
\sum
(\Psi_0^{\dagger })_{ik}
(\Psi_0)_{lj}
\left| f_i \right\ran
\left\lan {\cal U}(f_k) | {\cal U}(f_l) \right\ran
\left\lan f_j  \right| ,
\ena
where
\bea
 \label{VV}
\left\lan {\cal U}(f_k) | {\cal U}(f_l) \right\ran
=
\left\lan u_1(f_k) | u_1 (f_l)  \right\ran
+
\left\lan u_2(f_k)  | u_2 (f_l) \right\ran
+
\left\lan f_k | f_l \right\ran .
\ena
Let us rewrite the equation 
${\cal D}_z \left| {\cal U}(f_i) \right\ran = 0$
as
\bea
{\bf D}\,  {\bf u}({\bf f}_i ) = - {\bf f}_i ,
\ena
where
\bea
{\bf D} = 
\left(
\begin{array}{cc}
 B_2 - z_2 &  B_1 - z_1 \\
 - (B^{\dagger }_1 - \bar{z}_1) &  
  B^{\dagger }_2 - \bar{z}_2
\end{array}
\right) , \quad
{\bf u}({\bf f_i}) =
\left(
\begin{array}{c}
\left|  u_1 (f_i) \right\ran \\
\left|  u_2 (f_i) \right\ran
\end{array}
\right) , \quad
{\bf f}_i
=
\left(
\begin{array}{c}
\left| f_i \right\ran \, I \\
  0
\end{array}
\right).
\ena
Since the correspondence between the 
elements of ideal and 
vector zero-modes is one-to-one, 
we can consider the inverse operator of ${\bf D}$:
\bea
 {\bf u}({\bf f}_i) = - \frac{1}{{\bf D} } {\bf f}_i .
\ena
Then, (\ref{VV}) can be written
as
\bea
\label{VV2}
& &\left\lan {\cal U}(f_k) | {\cal U}(f_l) \right\ran \nn
& &=
{\bf u}^{\dagger}({\bf f}_k)\,
{\bf u}({\bf f}_l)
+
{\bf f}_k^{\dagger} {\bf f}_l
=
{\bf f}_k^{\dagger} 
\left(
\frac{1}{{\bf D}{\bf D}^{\dagger}}
+ 1
\right)
{\bf f}_l  \nn
& &=
\left\lan
f_k
\right|
I^{\dagger}\left(
\left(
\frac{1}{{\bf D}{\bf D}^{\dagger} }
\right)_{1\dot{1}} +1
\right)I
\left|
f_l
\right\ran ,
\ena
where
we denote the components of 
$({\bf D}{\bf D}^{\dagger} )^{-1}$ as
\bea
({\bf D}{\bf D}^{\dagger} )^{-1}
=
\left(
\begin{array}{cc}
({\bf D}{\bf D}^{\dagger} )^{-1}_{1\dot{1}} &
({\bf D}{\bf D}^{\dagger} )^{-1}_{1\dot{2}} \\
({\bf D}{\bf D}^{\dagger} )^{-1}_{2\dot{1}} &
({\bf D}{\bf D}^{\dagger} )^{-1}_{2\dot{2}} \\
\end{array}
\right).
\ena
From (\ref{VV2}), the matrix 
$C_{kl} = \left\lan {\cal U}(f_k) | {\cal U}(f_l) \right\ran $
has no zero-eigenvalue-vector 
and we can consider
$(C^{-1})_{kl}$.
The normalized minimal operator zero-mode is uniquely
determined (up to phase factor):
\bea
& &(\Psi_0)_{ij}
= (C^{-1/2})_{ij} . 
\ena


\section{Calculations
by the Method of 
Operator Symbols}\label{Nsymbol}

One can represent the 
equations over the algebra ${\cal A}_\zeta$
by mapping operators to
ordinary c-number functions
(operator symbols)
and using star product.
Some calculations become
simpler by the use of operator symbols.
The map from operators to
ordinary functions depends on
operator ordering procedures.
In order to express holes on D3-brane
(see section \ref{secIIB}),
we utilize normal symbol
which corresponds to the normal ordering.
Here we review this normal symbol.
For more detailed arguments on the
operator symbols,
see for example  \cite{20}
and references therein.
In this appendix, we use $\hat{\, \, \,}$ 
to denote the operators:
$\hat{x}^{\mu}$'s are noncommutative operators and
$x^{\mu}$'s are c-number coordinates of ${\bf R}^4$.

Let us consider normal ordered operator
of the form
\bea
 \label{Nop}
\hat{f}(\hat{x}) = 
\int \frac{d^4k}{(2\pi)^4} \, 
\tilde{f}(k) :  e^{ik\hat{x} } : \, ,
\ena
where $k\hat{x} := k_{\mu}\hat{x}^{\mu}$.
$: \, \, :$ denotes the normal ordering.
For the 
operator valued 
function 
(\ref{Nop}),
the corresponding {\bf normal symbol} 
is defined by
\bea
 \label{symb}
f_{\scriptscriptstyle{N}} (x) =
\int \frac{d^4k}{(2\pi)^4} \,
\tilde{f}(k) \ e^{ikx}\, , 
\ena
where $x^{\mu}$'s
are commuting coordinates of ${\bf R}^4$. 
We define $\Omega_N$ as a
map from operators to the normal symbols:
\bea
\Omega_N(\hat{f}(\hat{x}) ) = 
f_{\scriptscriptstyle N}(x)
:=
\int \frac{d^4k}{(2\pi )^4}
\left(
\left(2\pi \frac{\zeta}{4} \right)^2 
\mbox{Tr}_{\cal H} \left\{
\hat{f}(\hat{x})\ :e^{-ik\hat{x}}:
\right\} \right)
\,  e^{ikx}
\ena
Notice that from the relation
$\mbox{Tr}_{\cal H}
\bigl\{
: \exp \, ( ik\hat{x} ) : \bigr\}
= \left({2\pi}\frac{4}{\zeta} \right)^2 
   \delta^{(4)}(k)$, %
it follows
\bea
 \label{intr}
\left( 2\pi \frac{\zeta}{4}\right)^2
\mbox{Tr}_{\cal H}\, \hat{f} (\hat{x})
=
\int d^4x\,  
f_{\scriptscriptstyle{N}} (x).
\ena  
The inverse map of $\Omega_N$ is given by
\bea
& &\Omega_N^{-1}(f(x)) = 
\hat{f}^{\scriptscriptstyle{N}}(\hat{x})
:= 
\int \frac{d^4k}{(2\pi)^4}
\left(
\int d^4x f(x)\  e^{-ikx}
\right)\,
: e^{ik\hat{x}}: \, .
\ena
The {\bf star product}
of functions is defined by:
\bea
 \label{star}
f(x) \star_{\Omega_N} g(x) :=
\Omega_N (\Omega_N^{-1} (f(x)) \Omega_N^{-1} (g(x))) \, .
\ena
Since 
\bea
\label{starpro}
& &:e^{ik\hat{x}} :\, :e^{ik\hat{x}} :
=
e^{\bar{w}\hat{z}} e^{w \hat{\bar{z}} }
e^{\bar{w}' \hat{z} } e^{w'\hat{\bar{z}}}
=
e^{\frac{\zeta}{2} w\bar{w}'}
e^{(\bar{w} + \bar{w}') \hat{z} }
e^{(w + w' )\hat{\bar{z}} },
\ena
where
\bea
& &w_1  = -\frac{i}{2}(k_2 + i k_1) , \quad 
w_2  =  -\frac{i}{2} (k_4 + i k_3) , \nn
& & \bar{w}\hat{z}
 = \bar{w}_1 \hat{z}_1 + \bar{w}_2 \hat{z}_2 , 
\quad etc. \quad,
\ena
the explicit form of the star product is given by
\bea
f(z,\bar{z}) \star_{\Omega_N} g(z,\bar{z})
=
\left.
e^{\frac{\zeta}{2} 
\frac{\pa }{\pa \bar{z} }
\frac{\pa}{\pa {z'} }
}
f(z,\bar{z}) 
g(z',\bar{z}')
\right|_{ z' = z, \bar{z}'=\bar{z} }.
\ena
From the definition (\ref{star}), the star product
is associative:
\EQ
(f(x)\star_{\Omega_N} g(x) \, ) \star_{\Omega_N} h(x)
=
f(x)\star_{\Omega_N} (g(x) \star_{\Omega_N} h(x) \,).
\EN
If we use coherent states, the
expression of the
normal symbol becomes simpler.
The {\bf coherent states} 
$\left| \bar{z}_1 , \bar{z}_2 \right\ran$
are eigen states
of annihilation operators 
$\hat{\bar{z}}_1, \hat{\bar{z}}_2$:
\bea
\hat{\bar{z}}_1 \left| \bar{z}_1 , \bar{z}_2 \right\ran
&=& \bar{z}_1 
\left| \bar{z}_1 , \bar{z}_2 \right\ran \, ,  \nn
\hat{\bar{z}}_2
\left| \bar{z}_1 , \bar{z}_2 \right\ran
&=& \bar{z}_2 
\left| \bar{z}_1 , \bar{z}_2 \right\ran.
\ena
Then the normal symbol of operator
$\hat{f}$ is given by
\bea
 \label{fcohe}
f_{\scriptscriptstyle N} (z,\bar{z})
=
\left\lan \bar{z}_1 , \bar{z}_2 \right|
\hat{f}\left| \bar{z}_1 , \bar{z}_2 \right\ran.
\ena
(\ref{fcohe}) follows from (\ref{Nop}),(\ref{symb})
and
\bea
\left\lan \bar{z}_1 , \bar{z}_2 \right|
: e^{ik\hat{x}}:
\left| \bar{z}_1 , \bar{z}_2 \right\ran
&=&
\left\lan \bar{z}_1 , \bar{z}_2 \right|
e^{\bar{w}\hat{z}} e^{w \hat{\bar{z}} }
\left| \bar{z}_1 , \bar{z}_2 \right\ran
=
e^{\bar{w}z} e^{w \bar{z} } \nn
&=&
e^{ikx},
\ena
(we have normalized the coherent states as
$
\left\lan \bar{z}_1 , \bar{z}_2 \right.
\left| \bar{z}_1 , \bar{z}_2 \right\ran = 1
$).
From (\ref{fcohe}) it is easy to see
that the normal symbol 
$f_{\scriptscriptstyle N}(z,\bar{z})$
vanishes at
$(z_1,z_2)$ when the corresponding
operator $\hat{f}$ annihilates
$\left| \bar{z}_1 , \bar{z}_2 \right\ran$
or
$\left\lan \bar{z}_1 , \bar{z}_2 \right|$, i.e.
\bea
\hat{f}\left| \bar{z}_1 , \bar{z}_2 \right\ran
= 0 \, \, \, \mbox{or} \, \,
\left\lan \bar{z}_1 , \bar{z}_2 \right| \hat{f}
= 0 \quad  \Longrightarrow 
\quad f_{\scriptscriptstyle N}(z,\bar{z})
 = 0 .
\ena


\newpage
\newcommand{\NP}[1]{Nucl.\ Phys.\ {\bf #1}}
\newcommand{\AP}[1]{Ann.\ Phys.\ {\bf #1}}
\newcommand{\PL}[1]{Phys.\ Lett.\ {\bf #1}}
\newcommand{\CQG}[1]{Class. Quant. Gravity {\bf #1}}
\newcommand{\CMP}[1]{Comm.\ Math.\ Phys.\ {\bf #1}}
\newcommand{\PR}[1]{Phys.\ Rev.\ {\bf #1}}
\newcommand{\PRL}[1]{Phys.\ Rev.\ Lett.\ {\bf #1}}
\newcommand{\PRE}[1]{Phys.\ Rep.\ {\bf #1}}
\newcommand{\PTP}[1]{Prog.\ Theor.\ Phys.\ {\bf #1}}
\newcommand{\PTPS}[1]{Prog.\ Theor.\ Phys.\ Suppl.\ {\bf #1}}
\newcommand{\MPL}[1]{Mod.\ Phys.\ Lett.\ {\bf #1}}
\newcommand{\IJMP}[1]{Int.\ Jour.\ Mod.\ Phys.\ {\bf #1}}
\newcommand{\JHEP}[1]{J.\ High\ Energy\ Phys.\ {\bf #1}}
\newcommand{\JP}[1]{Jour.\ Phys.\ {\bf #1}}


\begin{thebibliography}{99}

\bibitem{CDS}
A. Connes, M. R. Douglas, A. Schwarz,
``Noncommutative Geometry and 
Matrix Theory: Compactification on Tori"
\JHEP{9802} (1998) 003.

\bibitem{BFSS}
T. Banks, W. Fischler, S. H. Shenker, L. Susskind,
``M Theory As A Matrix Model: A Conjecture"
\PR{D55} (1997) 5112.


\bibitem{IKKT}
N. Ishibashi, H. Kawai, Y. Kitazawa, A. Tsuchiya,
``A Large-N Reduced Model as Superstring"
\NP{B498} (1997) 467.


\bibitem{IKKTB}
H. Aoki, N. Ishibashi, S. Iso, H. Kawai, Y. Kitazawa, T. Tada
``Noncommutative Yang-Mills in IIB Matrix Model"
hep-th/9908141.

\bibitem{SI}
E. Witten,
``Small Instantons in String Theory"
\NP{B460} (1996) 541.

\bibitem{pinp4}
M. R. Douglas,
``Branes within Branes"
hep-th/9512077.


\bibitem{ADHMconst}
M. Atiyah, N. Hitchin, V. Drinfeld and Y. Manin, 
``Construction of Instantons"  
\PL{65A} (1978) 185. \\ 
E. Corrigan and P. Goddard, 
``Construction of instanton monopole solutions
and reciprocity"
\AP{154} (1984) 253. \\
S. Donaldson, 
``Instantons and Geometric
Invariant Theory"
\CMP{93} (1984) 453. 


\bibitem{iNakaj}
H. Nakajima, 
``Resolutions of moduli spaces of ideal instantons
 on ${\bf R}^4$ "
World Scientific, 1994, 129.

\bibitem{Nakaj}
H. Nakajima, 
``Heisenberg algebra and Hilbert schemes 
of points on projective surfaces",
Ann. of Math. {\bf 145}, (1997) 379, alg-geom/9507012; 
``Instantons and affine Lie algebra",
Nucl. Phys. Proc. Suppl. {\bf 46} (1996) 154, 
alg-geom/9510003.

\bibitem{LecNakaj}
H. Nakajima, 
``Lectures on Hilbert scheme of points on surfaces",
AMS University Lecture Series vol {\bf 18} (1999), 
ISBN: 0-8218-1956-9.


\bibitem{ABS}
O. Aharony, M. Berkooz, N. Seiberg,
``Light-Cone Description of (2,0) 
Superconformal Theories in Six Dimensions"
Adv. Theor. Math. Phys. {\bf 2} (1998) 119.

\bibitem{NS}
N. Nekrasov and A. Schwarz,
``Instantons on Noncommutative ${\bf R}^4$ 
  and $(2,0)$ Superconformal Six Dimensional
  Theory" \CMP{198} (1998) 689.

\bibitem{BN}
H. Braden, N. Nekrasov,
``Space-Time Foam From Non-Commutative 
Instantons"
hep-th/9912019.

\bibitem{Laza}
C. I. Lazaroiu,
``A noncommutative-geometric interpretation 
of the resolution of equivariant
instanton moduli spaces"
hep-th/9805132.


\bibitem{SW}
N. Seiberg and E. Witten,
``String Theory and Noncommutative Geometry"
\JHEP{9909} (1999) 032.


\bibitem{DQ}
M. R. Douglas, G. Moore,
``D-branes, Quivers, and ALE Instantons"
hep-th/9603167.

\bibitem{IonD}
C. Vafa,
``Instantons on D-branes"
\NP{B463} (1996) 435.

\bibitem{SumD}
H. Ooguri, C. Vafa
``Summing up D-Instantons"
\PRL{77} (1996) 3296.

\bibitem{DMVV}
R. Dijkgraaf, G. Moore, E. Verlinde, H. Verlinde,
``Elliptic Genera of Symmetric Products 
and Second Quantized Strings"
\CMP{185} (1997) 197.


\bibitem{Ours}
K. Furuuchi, H. Kunitomo and T. Nakatsu,
``Topological Field Theory and Second-Quantized Five-Branes"
\NP{B494} (1997) 144.

\bibitem{nakatsu}
T. Nakatsu,
``Toward Second-Quantization of D5-Brane"
\IJMP{A13} (1998) 923.


\bibitem{BPS}
J. A. Harvey, G. Moore,
``On the algebras of BPS states"
\CMP{197} (1998) 489.

\bibitem{Dbound}
G. Moore, N. Nekrasov, S. Shatashvili,
``D-particle bound states and generalized instantons"
hep-th/9803265.

\bibitem{KP}
V. A. Kazakov, I. K. Kostov, N. Nekrasov,
``D-particles, Matrix Integrals and KP hierachy"
hep-th/9810035.


\bibitem{HilbD}
A. Gorsky, N. Nekrasov, V. Rubtsov,
``Hilbert Schemes, Separated Variables, and D-Branes"
hep-th/9901089.

\bibitem{Matso}
Y. Matsuo,
``Matrix Theory, Hilbert Scheme and Integrable System"
\MPL{A13} (1998) 2731.


\bibitem{Sen}
A. Sen,
``Tachyon Condensation on the Brane Antibrane System"
\JHEP{9808} (1998) 012.

\bibitem{AH2}
H. Awata, S. Hirano, Y. Hyakutake,
``Tachyon Condensation 
and Graviton Production in Matrix Theory"
hep-th/9902158.

\bibitem{K1}
R. Minasian and  G. Moore,
``K-theory and Ramond-Ramond charge",
\JHEP{9711} (1997) 002.

\bibitem{K2}
E. Witten,
``D-Branes And K-Theory",
\JHEP{9812} (1998) 019.


\bibitem{Con}
A. Connes,
``Noncommutative Geometry"
Academic Press (1994).

\bibitem{Landi}
G. Landi,
``An Introduction to 
Noncommutative Space and their Geometry",
Lecture Notes in Physics: Monographs, m51
(Springer-Verlag, Berlin Heidelberg, 1997),
hep-th/9701078.

\bibitem{20}
D. Sternheimer,
``Deformation Quantization: Twenty Years After",
math/9809056. \\
C. Tzanakis and  A. Dimakis,
``On the uniqueness of the Moyal structure of 
phase-space functions"
q-alg/9605018.

\end{thebibliography}
\end{document}